\documentclass[runningheads]{llncs}
\usepackage[T1]{fontenc}
\usepackage{graphicx}
\usepackage{subcaption}
\usepackage{sidecap}
\usepackage{enumitem}
\usepackage{longtable}
\usepackage{mathtools}
\usepackage{amsfonts}
\usepackage{comment}

\usepackage{amssymb}
\makeatletter
\renewcommand*\env@matrix[1][*\c@MaxMatrixCols c]{%
	\hskip -\arraycolsep
	\let\@ifnextchar\new@ifnextchar
	\array{#1}}
\makeatother
\usepackage{booktabs}
\usepackage{mathrsfs}  
\usepackage{tikz}
\usetikzlibrary{positioning,calc,fit,arrows}
\usepackage{tkz-fct}
\usepackage{calc}
\usetikzlibrary{intersections}
\usetikzlibrary{decorations.shapes,shapes.geometric,shadows,arrows,automata,positioning,calendar,mindmap,backgrounds,scopes,chains,er,patterns,pgfplots.groupplots}
\tikzset{initial text={}, 
	double distance=2pt, 
	every state/.style = {draw = black, fill = grayfilling} 
}
\usetikzlibrary{calc}
\usepackage[framemethod=TikZ]{mdframed}

\usepackage{orcidlink}
\renewcommand{\orcidID}{\orcidlink}

\usepackage[outdir=./]{epstopdf}

\newcommand{\R}{\mathbb{R}}

\usepackage{xcolor}
\definecolor{lightblue}{rgb}{0.67, 0.9, 0.93}
\definecolor{lightgreen}{rgb}{0.67, 0.88, 0.69}
\definecolor{lightpink}{rgb}{1.0, 0.72, 0.77}
\definecolor{lightpurple}{rgb}{0.96, 0.73, 1.0}
\definecolor{lightyellow}{rgb}{0.98, 0.93, 0.37}
\definecolor{grayfilling}{gray}{0.95} 
\definecolor{grayshadow}{gray}{0.5} 
\colorlet{cherryred}{red!80!black}

\spnewtheorem{assumption}{Assumption}{\bfseries}{\itshape}
\spnewtheorem*{rexample}{Running example}{\itshape}{\itshape}
\newenvironment{prob}
{\vspace{.8em}\begin{mdframed}[backgroundcolor=white,shadow=true,shadowsize=5.5pt, shadowcolor=grayshadow]
\setlength{\belowdisplayskip}{0pt}%
\begin{problem}}{\end{problem}\vspace{.0em}\end{mdframed}\vspace{0em}}

\newcommand{\1}{\mathbf{1}} 
\newcommand{\norm}[1]{\left\lVert#1\right\rVert} 
\newcommand{\cdf}[1]{\mathrm{cdf}\left( #1 \right)}

\newcommand{\T}{^\top}

\newcommand{\xp}{x_+}
\newcommand{\xhp}{\hat x_+}

\newcommand{\cdotx}{\,\cdot\,}

\ifdefined\U
\renewcommand{\U}{\mathbb{U}}
\else
\newcommand{\U}{\mathbb{U}}
\fi

\newcommand{\C}{{\mathbf{{C}}}}

\newcommand{\M}{{\mathbf{{M}}}}
\newcommand{\Mh}{\widehat{\M}}

\newcommand{\agent}{{\mathbf{{A}}}} 
\newcommand{\env}{{\mathbf{{E}}}} 
\newcommand{\agentred}{{\agent_{r\!}}} 
\let\ORGmathfrak\mathfrak
\def\mathfrak#1{{\ifnum\pdfstrcmp{#1}{S}=0 \mathfrakS\else\ORGmathfrak{#1}\fi}}
\newcommand{\mathfrakS}{\mathpalette\bigmathfrakS\relax}
\newcommand{\bigmathfrakS}[2]{\scalebox{1.6}{$#1\ORGmathfrak{s}$}}
\newcommand{\simulator}{{\mathbf{{S}}}} 
\newcommand{\gua}{{\mathbf{{G}}}} 
\newcommand{\adv}{{\mathfrak{{n}}}} 
\newcommand{\advAmbSet}{{\mathfrak{{N}}}} 
\newcommand{\Uh}{\hat{\mathbb{U}}}

\newcommand{\D}{\mathbb{D}}
\newcommand{\Dh}{\hat{\D}}
\renewcommand{\O}{\mathbb{O}}
\newcommand{\Oh}{\hat{\O}}

\newcommand{\satisfies}{\vDash}

\newcommand{\Tr}{\mathbf{t}}
\newcommand{\Trh}{\hat{\mathbf{t}}}

\newcommand{\X}{\mathbb{X}}
\newcommand{\Xh}{\hat{\mathbb{X}}}

\usepackage{transparent}

\newcommand{\rel}{\mathcal{R}}

\newcommand{\CA}[1]{\mathsf{#1}}

\newcommand{\AP}{\mathsf{AP}}

\newcommand{\notltl}{\neg}
\newcommand{\andltl}{\wedge}

\newcommand{\Next}{\ensuremath{\bigcirc}}

\newcommand{\Until}{\mathbin{\CA{U}}}

\newcommand{\alphabeth}{\Sigma}
\newcommand{\word}{\boldsymbol{\omega}}

\newcommand{\letter}{l}
\newcommand{\True}{\operatorname{\mathsf{true}}}

\newcommand{\N}{\mathbb{N}}
\newcommand{\normal}{\mathcal{N}} 
\usepackage{accents}

\newcommand{\borel}[1]{\mathcal{B}(#1)}

\newcommand{\InFu}{\mathbf{i}_{u}} 

\newcommand{\h}{h}
\newcommand{\Y}{\mathbb{Y}}
\newcommand{\Yh}{\hat{\mathbb{Y}}}

\newcommand{\dist}{\mathbf{d}_{\Y}}

\newcommand{\W}{\mathbb{W}}
\newcommand{\Wt}{\boldsymbol{v}}

\newcommand{\simrel}[4]{#1\hspace{-3pt}{\texttransparent{0}\preceq}\hspace{-1pt}{\preceq^{#3}_{#4}}\,#2} 
\newcommand{\simreltwo}[4]{#2\sqsubseteq^{#4}_{#3}#1} 
\usepackage{xspace}

\renewcommand{\P}{\mathbb{P}}

\newcommand{\new}[1]{{#1}}

\graphicspath{{./Figures/}} 

\usepackage{import}
\usepackage{xifthen}
\usepackage{pdfpages}
\usepackage{transparent}

\makeatletter
\renewcommand\paragraph{\@startsection{paragraph}{4}{\z@}%
                       {-12\p@ \@plus -4\p@ \@minus -4\p@}%
                       {-0.5em \@plus -0.22em \@minus -0.1em}%
                       {\normalfont\normalsize\bfseries}}
\makeatother

\newif\iflong
\longtrue

\begin{document}
\title{
Formal Control for Uncertain Systems via Contract-Based Probabilistic Surrogates
\iflong
(Extended Version)
\fi
}
\titlerunning{Formal Contract-Based Control for Uncertain Systems}
%
\author{Oliver Sch\"on\inst{1}\orcidID{0000-0002-0214-6455} \and Sofie Haesaert\inst{2}\orcidID{0000-0003-4749-4688} \and Sadegh Soudjani\inst{3}\orcidID{0000-0003-1922-6678}}
\authorrunning{O. Sch\"on et al.}
%
\institute{Newcastle University, 
Newcastle upon Tyne, 
United Kingdom \email{o.schoen2@ncl.ac.uk}\and 
Eindhoven University of Technology,
Eindhoven,
The Netherlands \email{s.haesaert@tue.nl}\and
Max Planck Institute for Software Systems,
Kaiserslautern, 
Germany \email{sadegh@mpi-sws.org}}
\maketitle 
\begin{abstract} 
The requirement for identifying accurate system representations has not only been a challenge to fulfill, but it has compromised the scalability of formal methods, as the resulting models are often too complex for effective decision making with formal correctness and performance guarantees.
Focusing on probabilistic simulation relations and surrogate models of stochastic systems, we propose an approach that significantly enhances the scalability and practical applicability of such simulation relations by eliminating the need to compute error bounds directly.
As a result, we provide an abstraction-based technique that scales effectively to higher dimensions while addressing complex nonlinear agent-environment interactions with infinite-horizon temporal logic guarantees
amidst uncertainty.
Our approach trades scalability for conservatism favorably, as demonstrated on a complex high-dimensional vehicle intersection case study.
\end{abstract}
\begin{figure}
    \def\svgwidth{1\linewidth}
    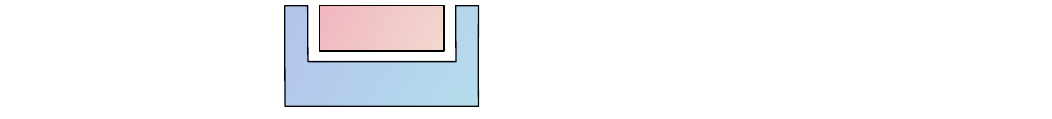

  \setlength{\abovedisplayskip}{0pt}
    \setlength{\belowdisplayskip}{0pt}
    \setlength{\abovecaptionskip}{0pt}
    \setlength{\belowcaptionskip}{0pt}
  \caption{
  Design a robust controller $\C$ for an agent $\agent$ based on finitely many surrogate models $\{\simulator_i\}_{i\in\mathcal{I}}$ and obtain formal guarantees for its performance (with respect to some specification $\psi$ and satisfaction probability $p_{\text{\normalfont{c}}}$) once deployed in the concrete environment $\env$. The model discrepancies arising from uncertainty and incomplete information
  are modeled
  via an uncertain parametrization $\theta\in\Theta$ and a nondeterministic adversary $\adv\in\advAmbSet$.
  }
  \label{fig:idea}
\end{figure}
\setcounter{footnote}{0} 

\section{Introduction}
To pave the way for the safe and responsible adoption of autonomous agents --- especially in high-risk applications such as \emph{cyber-physical systems} (CPSs) --- the creation of formal techniques for decision making and control is proving to be a key puzzle piece \cite{annaswamy2024control,yohsua2024international,dalrymple2024towards,luckcuck2023using,provan2024formal}.
In the pursuit of developing control software for agents that implements specifications with a certified probability, a main task is establishing a formal connection between the concrete environment $\env$ (see Fig.~\ref{fig:idea}) in which these agents operate and (certification) results obtained via mathematical models (e.g., via surrogate models $\{\simulator_i\}_{i\in\mathcal{I}}$).
Available approaches are challenged by the uncertain nature of actual CPS dynamics and model imperfections, as indicated in Fig.~\ref{fig:idea}, respectively, by an uncertain parametrization $\theta\in\Theta$ and an adversary\footnote{The adversary $\adv$ --- in the literature sometimes referred to as \emph{nature} --- compensates for model inadequacies by disturbing the system. We express its nondeterministic nature via the ambiguity set $\advAmbSet$.} $\adv\in\advAmbSet$.
The resulting \emph{environmental ambiguity} renders a strict sampling-based cover of all possible scenarios $\{\simulator_i(\theta,\adv)| i\in\mathcal{I},\theta\in\Theta,\adv\in\advAmbSet\}$
straight-out impossible.

Rooted in the control systems community, formal control synthesis approaches have been developed with the theoretical foundations of dynamical systems such as CPSs in mind \cite{belta2016formal,lavaei2021automated,yin2024survey}.
For instance, \emph{(stochastic) simulation relations} \cite{segala1995probabilistic,BK08,haesaert2017verification,haesaert2021simrels} are designed to bridge the gap between discrete-state models and their uncountable counterparts.
This includes transferring guarantees that quantify the risk of undesired behavior even when the system behaves nondeterministically and available models are incomplete or imprecise~\cite{schon2024bayesian}. 

\begin{SCfigure}[][tbh]
    \def\svgwidth{.5\linewidth}
\begingroup%
  \makeatletter%
  \providecommand\color[2][]{%
    \errmessage{(Inkscape) Color is used for the text in Inkscape, but the package 'color.sty' is not loaded}%
    \renewcommand\color[2][]{}%
  }%
  \providecommand\transparent[1]{%
    \errmessage{(Inkscape) Transparency is used (non-zero) for the text in Inkscape, but the package 'transparent.sty' is not loaded}%
    \renewcommand\transparent[1]{}%
  }%
  \providecommand\rotatebox[2]{#2}%
  \newcommand*\fsize{\dimexpr\f@size pt\relax}%
  \newcommand*\lineheight[1]{\fontsize{\fsize}{#1\fsize}\selectfont}%
  \ifx\svgwidth\undefined%
    \setlength{\unitlength}{2593.6842041bp}%
    \ifx\svgscale\undefined%
      \relax%
    \else%
      \setlength{\unitlength}{\unitlength * \real{\svgscale}}%
    \fi%
  \else%
    \setlength{\unitlength}{\svgwidth}%
  \fi%
  \global\let\svgwidth\undefined%
  \global\let\svgscale\undefined%
  \makeatother%
  \begin{picture}(1,1.22024382)%
    \lineheight{1}%
    \setlength\tabcolsep{0pt}%
    \put(0,0){\includegraphics[width=\unitlength,page=1]{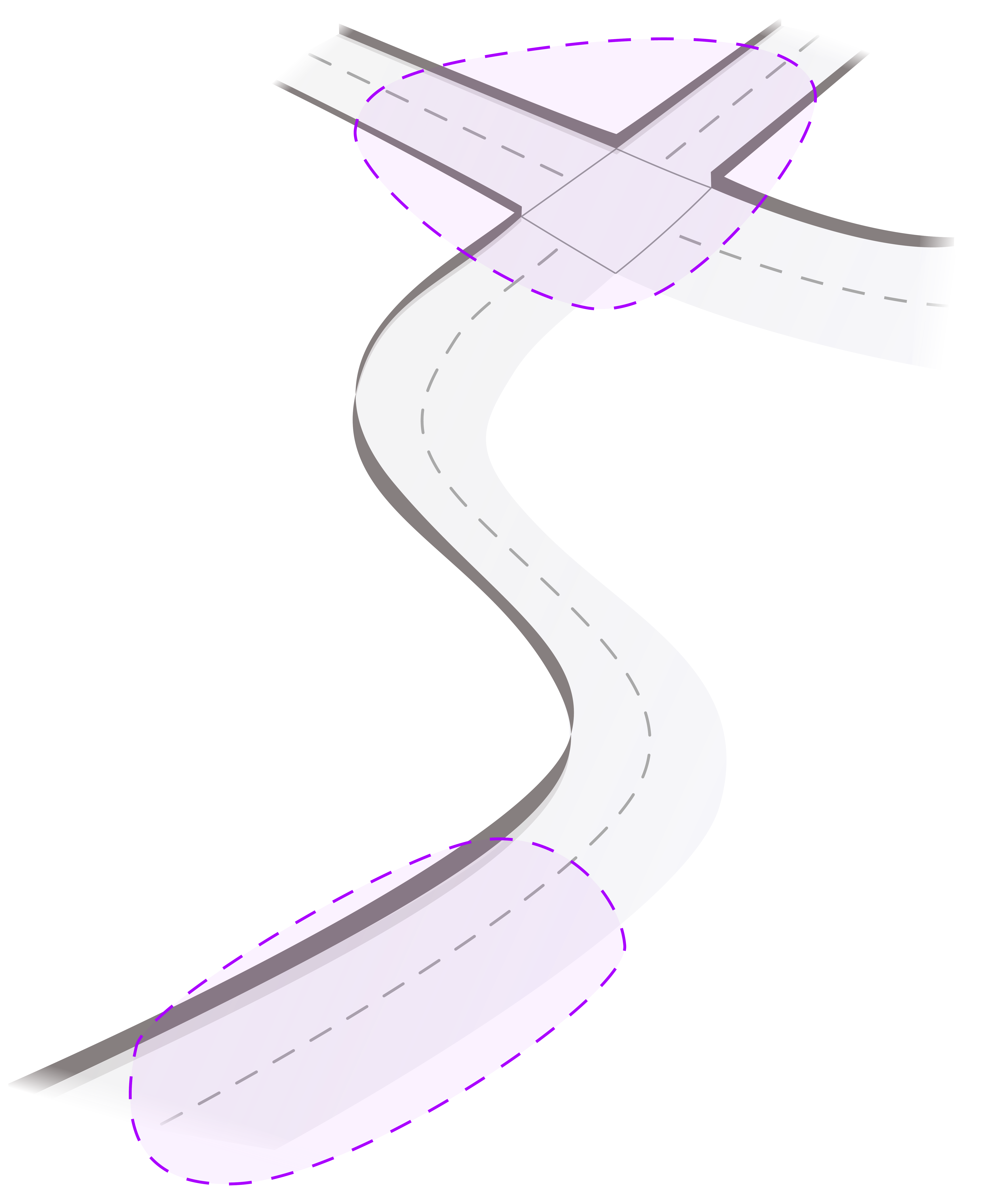}}%
    \put(0.49052494,1.07){\color[rgb]{0,0,0}\makebox(0,0)[t]{\lineheight{1.25}\smash{\begin{tabular}[t]{c}$\env_2$\end{tabular}}}}%
    \put(0.52,0.30369674){\color[rgb]{0,0,0}\makebox(0,0)[t]{\lineheight{1.25}\smash{\begin{tabular}[t]{c}$\env_1$\end{tabular}}}}%
    \put(0.30237498,0.085){\color[rgb]{0,0,0}\makebox(0,0)[t]{\lineheight{1.25}\smash{\begin{tabular}[t]{c}$\agent$\end{tabular}}}}%
    \put(0.685,1.06){\color[rgb]{0,0,0}\makebox(0,0)[lt]{\lineheight{1.25}\smash{\begin{tabular}[t]{l}$\agent$\end{tabular}}}}%
    \put(0,0){\includegraphics[width=\unitlength,page=2]{situations_render.pdf}}%
  \end{picture}%
\endgroup%

    \caption{Situational awareness: The environment can be broken down into situations $\env_1,\env_2,\ldots$ that individually describe its interaction with the agent $\agent$ in a more concrete context.
    Note that, for the depicted interactions $\agent\times\env_1$ and $\agent\times\env_2$, models describing the longitudinal dynamics of the vehicles may be sufficient for effective decision making.
    }
    \label{fig:situations}
\end{SCfigure}

A key challenge for formal methods such as abstraction-based approaches, however, is their limited scalability.
CPSs are complex, high-dimensional systems and the semantics required to express their desired behavior can often only be handled over finite abstractions, which partition the underlying state space with a computational complexity exponential in the dimension of the system.
To improve scalability, multiple studies have been conducted including the use of adaptive partitioning, \emph{model-order reduction} (MOR), and compositional techniques \cite{esmaeil2013adaptive,banse2023data,chou2017study,alur2001compositional,hahn2013compositional,lavaei2022compositional,wang2025unraveling}.
These studies require appropriate assumptions, ranging from continuity to
small-gain, input-to-state stability, bounded subsystem interaction, and decomposable specifications.
By using the natural split between environment and agent in autonomous systems, we bypass these assumptions by leveraging the \emph{situational awareness} (SA) block that is an integral part of the architecture of many modern autonomous systems~\cite{casablanca2024symaware}.
This includes identifying which situation $\env_1,\env_2,\ldots$ the agent is in and provisioning a sufficiently rich situation-dependent model to aid in effective decision making.
Consider the situations shown in Fig.~\ref{fig:situations}.
In essence, the SA block of an agent $\agent$ processes (sensor) information, equipping it with a sufficient uncertainty/risk-aware understanding of its environment
to make informed decisions.
In this paper, we will mainly focus on the decision making/control synthesis part, assuming access to a situation supplied by the SA block.
In the following, we assume that for each situation $i=1,2,\ldots$ the emanating interaction of the agent and the environment is entirely captured by the interconnected system $\agent\times\env_i$.
As an example, an autonomous vehicle (the agent $\agent$) may encounter situations such as $\env_1$ and $\env_2$ illustrated in Fig.~\ref{fig:situations}.
Even though each system $\env_i$ may still be complex and high-dimensional, we will assume that the situations are defined sufficiently elemental such that for every situation $i=1,2,\ldots$ the encapsulated dynamics \emph{relevant} to the interaction $\agent\times\env_i$ and the specification $\psi$ can be distilled into a much simpler form.
For instance, following a car at a safe distance (see $\agent\times\env_1$ in Fig.~\ref{fig:situations}) concerns mainly the systems' longitudinal dynamics.
Consequently, we abandon the family of potentially complex and high-dimensional environments $\env_1,\env_2,\ldots$ and replace it with a set of much simpler surrogate models $\simulator_1,\simulator_2,\ldots$ as identified by the agent's SA block~\cite{casablanca2024symaware}.
For simplicity, we will call these surrogate models \emph{simulators}.
In the remainder of this work, we will focus our attention on situations in isolation by assuming that the system is in a single situation at a given time. Thus, we drop the subscript $i$ and simply refer to $\env$ as the environment and $\simulator$ as the corresponding simulator.
With this in mind, we make the following contributions. 

\smallskip
\noindent\textbf{Contributions.}
(1) We formalize a type of system interrelation called \emph{behavioral inclusion} for stochastic systems \emph{subject to model uncertainty}, that allows us to replace complex models with simpler surrogates by introducing an adversary.
(2) We extend the definition of sub-simulation relations to systems subject to nondeterministic adversaries, by compensating the induced ambiguity in the stochastic coupling.
(3) Based on the aforementioned two contributions, we establish an \emph{assume-guarantee contract} (AGC) that allows us to reason about $\agent\times\env$ via $\agentred\times\simulator$ as a proxy. As a result, we design robust controllers via the monolithic system $\agentred\times\simulator$ that can be readily deployed to $\agent\times\env$.
For simplicity, we limit ourselves to nonlinear systems with additive Gaussian noise.
(4) We demonstrate the power of the proposed AGC framework on a high-dimensional traffic intersection case study. In particular, we use the new result to synthesize a controller for a 9-dimensional nonlinear stochastic system and a complex monolithic specification via a 4-dimensional linear proxy, resulting in a reduction in the number of required partitions by orders of magnitude.
Notably, we emphasize that these contributions are for systems with inherent \emph{model uncertainty}.

\medskip

The remainder of this manuscript is organized as follows. 
After reviewing the related work in Section~\ref{app:related_work}, we present in Section~\ref{sec:preliminaries} the general notation and the problem setup. In Section~\ref{sec:relations}, we formalize the types of system interrelations used to provide the main assume-guarantee result in Section~\ref{sec:agc}.
We break down the steps necessary to establish the relations to design a controller along with robust guarantees that cover both environment and system in Section~\ref{sec:simtoreal}.
The case study in Section~\ref{sec:experiments} demonstrates the efficacy of the proposed AGC approach in obtaining results on a complex composed system.
Section~\ref{sec:conclusion} concludes the paper.
Due to space constraints, 
we defer the proofs to 
\iflong Appendix~\ref{app:proof_proxy_theorem}.
\else an extended version~\cite{extendedVersion}.
\fi

\section{Related Work}\label{app:related_work}
The advancement of formal control synthesis increasingly targets scalable frameworks that overcome com\-plex en\-vironmental inter\-actions and un\-certain\-ties through pivotal strategies such as assume–guarantee frameworks and compositional MOR. Although abstraction–free methods using barrier certificates are effective for low–level control, their extension to high–level decision making is challenged by temporal logic complexities and the need to balance accuracy with efficiency --- thereby prompting reliance on abstraction–based approaches \cite{matni2024controlarchitecture,nayak2023context}.

In networked systems, many techniques assume that networks consist of weakly interconnected sub–components with local sub–specifications, as illustrated in \cite{mallik2018compositional} using disturbance bisimulation and by \cite{kazemi2024assume} with reinforcement learning in an AGC framework; meanwhile, abstraction–free AGC methods have mainly focused on linear systems, where \cite{ghasemi2020compositional} address linear time–variant systems via convex parametric contracts and \cite{nuzzo2019stochastic} transform bounded stochastic STL contracts into mixed integer programs with chance constraints.
%

Modeling real–world systems with probabilistic behavior introduces significant uncertainty, which many approaches tackle using variants of \emph{robust Markov decision processes} (RMDPs) --- including interval and parametric MDPs to robustify against epistemic uncertainty \cite{puterman2014markov,schon2024bayesian,Badings2022epistemicUncert,hahn2019interval} --- though high–dimensional extensions remain challenging. For instance, while simulation–relation–based synthesis via parametric MDPs with MOR error quantification has been studied \cite{VanHuijgevoort2020SimQuant,schon2024bayesian}, no known IMDP–based MOR approaches exist; similarly, distributionally robust RMDPs based on ambiguity sets are computationally feasible only under fixed deterministic dynamics \cite{gracia2023distributionally}, with MOR for linear RMDPs explored in \cite{pulch202010}.
Similar to our work, integration of antagonistic non–determinism with stochastic simulation relations was pioneered in \cite{Zhong2023game}.

Abstraction–free methods that propagate reachable sets are often limited to linear systems \cite{arcari2023stochastic}. Recent work by \cite{jafarpour2024probabilistic} decouples deterministic and stochastic influences to generate reachable tubes via a deterministic proxy --- an idea we mirror by replacing a rich stochastic system with a less complex nondeterministic proxy. 
In a similar vein, \cite{coppola2024enhancing} relate data from the original stochastic system to a nondeterministic transition system over probability measures using the scenario approach, while \cite{Kwiatkowska2010AGC} focus on AGCs for verifying discrete–state systems exhibiting both probabilistic and nondeterministic behaviors. Complementary to our work --- which assumes perfect perception --- \cite{astorga2023perception} establish contracts on environmental estimates from deterministic neural perception components, with \cite{puasuareanu2023closed} considering a probabilistic variant.

\section{Preliminaries and Problem Statement}\label{sec:preliminaries}
\paragraph{Notation.}
The transpose of a matrix $A$ is denoted by $A\T$. We write $I_n\in\R^{n\times n}$ for the identity matrix and $0_{n\times m}\in\R^{n\times m}$ for the zero matrix. 
Let $(\X,\borel{\X})$ be a \emph{Borel measurable space}, with a \emph{Polish} sample space $\X$ and a Borel $\sigma$-algebra $\borel{\X}$~\cite{bogachev2007measure}.
A \emph{measure} $\nu:\borel{\X}\to\R_{\ge0}$
is a \emph{probability measure} if $\nu(\X)=1$ and a \emph{sub-probability measure} if $\nu(\X)\le1$. A \emph{probability space} is given by $(\X,\borel{\X},p)$, with realizations $x\sim p$.
For two measurable spaces $(\X,\borel{\X})$ and $(\Y,\borel{\Y})$, a \emph{\mbox{(sub-)}probability kernel} is a mapping $\mathbf{p}:\X\times\borel{\Y}\to\R_{\ge0}$ such that for each $x\in\X$, $\mathbf{p}(x,\cdotx)$ is a measure on $(\Y,\borel{\Y})$ and for each $B\in\borel{\Y}$ the function $x\mapsto \mathbf{p}(x,B)$ is measurable; we write $\mathbf{p}(\cdotx|x)$ for the measure associated to $x$, which is a (sub-)probability measure when $\mathbf{p}(x,\cdotx)$ is. 
The \emph{Dirac delta} measure $\delta_a:\borel{\X}\to\{0,1\}$ at $a\in\X$ is defined by $\delta_a(A)=1$ if $a\in A$ and $\delta_a(A)=0$ otherwise. 
The Gaussian measure with mean $\mu\in\R^n$ and covariance matrix $\Sigma\in\R^{n\times n}$ is defined by
$\normal(dx|\mu, \Sigma) := dx/(\sqrt{(2\pi)^n \left|\Sigma\right| })\exp(-\frac{1}{2}(x-\mu)\T\Sigma^{-1}(x-\mu)),$
with $|\Sigma|$ being the determinant of $\Sigma$. 
We denote the Gaussian measure truncated to a support set $A\in\borel{\X}$ as 
$\normal_A(dx|\mu,\Sigma) := \1_A(x)\,\normal(dx|\mu,\Sigma)/\int_{A}\normal(d\xi|\mu,\Sigma),$
where the indicator function $\1_A(x)$ of a measurable set $A\in\borel{\X}$ evaluates to $\1_A(x)=1$ if $x\in A$ and $\1_A(x)=0$ otherwise.
The \emph{cumulative distribution function} (CDF) of a Gaussian distribution is
$\smash{\cdf{x} := \int_{-\infty}^x \frac{1}{\sqrt{2\pi}} \exp(-\xi^2/2)\;d\xi.}$
Given sets $A,B$, a relation $\rel\subset A\times B$ relates $x\in A$ and $y\in B$ if $(x,y)\in\rel$. A metric on $\Y$ is a function $\dist:\Y\times\Y\to\R_{\ge0}$ satisfying $\dist(y_1,y_2)=0$ iff $y_1=y_2$, $\dist(y_1,y_2)=\dist(y_2,y_1)$, and $\dist(y_1,y_3)\le \dist(y_1,y_2)+\dist(y_2,y_3)$ for all $y_1,y_2,y_3\in\Y$.
An example is the norm function $\mathbf{d}_\Y(y_1,y_2):=\norm{y_1-y_2}$.
Given two functions $f,g$ with suitable (co-)domains, we indicate their composition as $f\circ g:=f(g(\cdotx))$.

\subsection{Game Setup}
We consider the interaction between an agent $\agent$ and a complex environment $\env$, where $\env$ is elemental.
In essence, this means that the dynamics of $\env$ \emph{relevant} to the considered situation (as defined below by virtue of the domain of $\env$) can be distilled into a small subset of equations.

We cast the interaction as a $2\frac{1}{2}$-player game, 
where the two main players $\agent$ and $\env$ are parameterized discrete-time nonlinear dynamical systems with uncertain parameters $\theta^\agent\in\Theta^\agent$ and $\theta^\env\in\Theta^\env$, respectively, subject to additive stochastic noise (the remaining $\frac{1}{2}$-player), i.e., systems of the form
\begin{align}
    \agent(\theta^\agent):&\left\{\begin{array}{ll}
        x^\agent_{t+1} &= f^\agent(x^\agent_t, o^\agent_t, u^\agent_t; \theta^\agent) + w^\agent_t\\[.2em]
	y^\agent_t &= h^\agent(x^\agent_t)
    \end{array}\right., && w^\agent_t \sim \mathcal{N}(\cdotx|0,\Sigma^\agent)\label{eq:agent},\\[.4em]
    \env(\theta^\env):&\left\{\begin{array}{ll}
        x^\env_{t+1} &= f^\env(x^\env_t, o^\env_t; \theta^\env) + w^\env_t\\[.2em]
	y^\env_t &= h^\env(x^\env_t)
    \end{array}\right., && w^\env_t \sim \mathcal{N}_{\W^\env\!}(\cdotx|0,\Sigma^\env),\label{eq:environment}
\end{align}
where the state, input, observation, noise, and output of system $\M\in\{\agent,\env\}$ at the $t^{\text{th}}$ time step are denoted by $x^\M_t\in\X^\M$, $u^\agent_t\in\U^\agent$, $o^\M_t\in\O^\M$, $w^\M_t\in\W^\M$, and $y^\M_t\in\Y^\M$, respectively. 
The functions $f^\M$ and $h^\M$ specify, respectively, the parameterized state evolution of the system and the observation map. The process noise $w^\M_t\in\W^\M$ constitutes an independent, identically distributed (i.i.d.) noise sequence.
Whilst the noise of the agent $\agent$ can be unbounded (and even from an arbitrary, potentially $\theta^\agent$-parametrized continuous distribution as shown in the paper \cite{Schoen2023GMM}), we raise the following assumption on $\env$.
\begin{assumption}[Boundedness]\label{asm:boundednoise}
Let $\W^\env$ and $\X^\env$ be bounded.
\end{assumption}
In this paper, we assume bounded noise to over-approximate the behavior of $\env$ in a simpler surrogate model.
Furthermore, without loss of generality, we consider autonomous environments, i.e., $\U^\env:=\emptyset$.
Our task is to design a state-feedback controller $\C$ for the agent $\agent$, that is
\begin{align*}
    \C:\X^\agent\times\X^\env\rightarrow\U^\agent.    
\end{align*}
The systems $\agent$ and $\env$ are interconnected via the operator $\times:\X^\agent\times\X^\env\rightarrow\O^\env\times\O^\agent$, written as $\agent\times\env$.
Unless specified otherwise, we assume $\times$ is constructed with $o^\env = x^\agent$ and $o^\agent = x^\env$.
Throughout the paper, we use the following running example.
\begin{rexample}
Consider an interaction between two vehicles $\agent$ and $\env$. We assume that the dynamics of the agent are given by the stochastic linear system
\begin{align}
    \agent(\theta^\agent): \left\lbrace
    \begin{array}{ll}	
        \begin{bmatrix}
            \xi^\agent_{t+1}\\
            v^\agent_{t+1}\\
            s^\agent_{t+1}
        \end{bmatrix}
        &= \begin{bmatrix}
            a_1&0&0\\
            a_2&\theta^\agent&0\\
            a_3&\tau&1
	\end{bmatrix}x_t^\agent 
        + \begin{bmatrix}
	    b\\\tau\\0
	\end{bmatrix}u_t^\agent + w^\agent_t\\[1.6em]
    y_t^\agent &= [v^\agent_t,s^\agent_t]\T
    \end{array}
    \right.,\label{eq:agent_car}
\end{align}
with state $x_t^\agent:=\smash{[\xi^\agent_{t},v^\agent_{t},s^\agent_{t}]\T\in\X^\agent}$, control input $u_t^\agent\in\U^\agent$, process noise $w^\agent_t\sim 
p^\agent_w(\cdotx):=
\smash{p^\agent_{w,\xi}p^\agent_{w,v}p^\agent_{w,s}}$, time discretization $\tau=0.5$, fixed parameters $a_1,a_2,a_3,b \in \R$, and an uncertain parameter $\theta^\agent\in\Theta^\agent:=[0.79, 0.81]$.
\iflong 
We defer further details to Appendix~\ref{app:agent_dynamics}.
\else We defer further details to \cite[Appendix~C.1]{extendedVersion}.
\fi
The ambiguity introduced by the stochastic noise $w^\agent_t$ and the uncertain parameter $\theta^\agent$ may arise from unmodeled dynamics, friction, aerodynamic drag, driver behavior, etc.

We assume that the dynamics of the environment are described by an autonomous six-dimensional nonlinear stochastic system
\begin{equation}
    \env(\theta^\env): \left\lbrace
    \begin{array}{ll}	
        \begin{bmatrix}
            \beta_{t+1}^\env\\
            \mathrm{d}\Psi_{t+1}^\env\\
            \Psi_{t+1}^\env\\
            v_{t+1}^\env\\
            s_{x,t+1}^\env\\
            s_{y,t+1}^\env
        \end{bmatrix} &=
    \begin{bmatrix}
		\beta_t^\env + \tau \dot\beta(x_t^\env)\\
		\mathrm{d}\Psi_{t}^\env + \tau \ddot\Psi(x_t^\env)\\
            \Psi_{t}^\env + \tau \mathrm{d}\Psi_{t}^\env\\
            \theta^\env v_{t}^\env\\
            s_{x,t}^\env + \tau v_t^\env\cos(\beta_t^\env+\Psi_t^\env)\\
            s_{y,t}^\env + \tau v_t^\env\sin(\beta_t^\env+\Psi_t^\env)
	\end{bmatrix}        
        + \begin{bmatrix}
	        0&0&0\\
                0&0&0\\
                0&0&0\\
                a_4&0&0\\
                a_5&0&0\\
                0&0&0
	\end{bmatrix}o_t^\env 
        + w_t^\env\\[3.6em]
    y_t^\env &= [v_t^\env, s_{x,t}^\env]\T
    \end{array}
    \right.,\label{eq:env_car}
\end{equation}
characterized by a state $x_t^\env:=[\beta_{t}^\env,\mathrm{d}\Psi_{t}^\env,\Psi_{t}^\env,v_{t}^\env,s_{x,t}^\env,s_{y,t}^\env]\T\in\X^\env$,
observation $o_t^\env\in\X^\agent$,
fixed parameters $a_4,a_5 \in \R$,
uncertain parameter $\theta^\env\in\Theta^\env:=[0.79,0.81]$,
and derivative terms $\dot\beta(x_t^\env)$ and $\ddot\Psi(x_t^\env)$ provided in 
\iflong
Appendix~\ref{app:env_dynamics}.
\else
\cite[Appendix~C.2]{extendedVersion}.
\fi
As alluded to in Assumption~\ref{asm:boundednoise}, the environment's process noise $w_t^\env\sim p^\env_w(\cdotx):=p^\env_{w,\beta}\cdots p^\env_{w,s_y}$ is constricted to a bounded support $\W^\env:=\W^\env_{\beta}\times\cdots\times\W^\env_{s_y}$.
\end{rexample}

For brevity, the resulting composed system parametrized by $\theta:=(\theta^\agent,\theta^\env)$ can be written as $(\agent\times\env)(\theta):=\agent(\theta^\agent)\times\env(\theta^\env)$. Furthermore, in the remainder, we may drop the explicit parametrization for conciseness.
With this, the game $(\agent\times\env)(\theta)$ unfolds as follows:
In every iteration $t\in\N$ of the game, the agent $\agent$ and environment $\env$ start from some initial states $x_t^\agent,x_t^\env$.
The controller $\C:\X^\agent\times\X^\env\rightarrow\U^\agent$ chooses a control input $u^\agent_t\in\U^\agent$.
Based on $u_t^\agent$ and the observation $o_t^\agent$, the agent system $\agent$ evolves to a successor state $x_{t+1}^\agent=f^\agent(x^\agent_t, o_t^\agent, u^\agent_t) + w^\agent_t$ by eliciting a noise realization $w^\agent_t\sim \mathcal{N}(\cdotx|0,\Sigma^\agent)$.
Simultaneously, the environment $\env$ evolves to $x_{t+1}^\env=f^\env(x^\env_t, x^\agent_t) + w^\env_t$ with $w^\env_t\sim \mathcal{N}_{\W^\env}(\cdotx|0,\Sigma^\env)$. 
In every iteration $t\in\N$, the systems emit outputs $y_t^\agent=h^\agent(x_t^\agent)$ and $y_t^\env=h^\env(x_t^\env)$, which will be used to determine the behavior of the systems according to specifications defined in Subsection~\ref{sec:specifications}. The setup can be generalized by defining a shared output mapping over the product space $\X^\agent\times\X^\env$.

\subsection{Temporal Logic Specifications}\label{sec:specifications}
Consider a set of atomic propositions $AP := \{ P_1, \ldots, P_L \}$ defining an \emph{alphabet} $\alphabeth := 2^{AP}$, where any \emph{letter} $\letter\in\alphabeth$ is composed of a set of atomic propositions.
An infinite string of letters forms a \emph{word} $\word=\letter_0\letter_1\letter_2\ldots\in\alphabeth^{\mathbb{N}}$.
Specifications imposed on the behavior of the system are defined as formulas composed of atomic propositions and operators. We consider the \emph{co-safe} subset of \emph{linear-time temporal logic} properties (scLTL) \cite{kupferman2001model}, for which violation can be determined from a finite bad prefix of a word. scLTL formulas are constructed according to the following syntax:
\begin{equation*}
    \psi ::=  \True\ |\  P \ |\ \notltl P\ |\ \psi_1 \vee\psi_2  \ |\ \psi_1 \andltl \psi_2 \ |\ \psi_1 \Until \psi_2 \ |\ \Next \psi,
\end{equation*}
where $P\in \AP$ is an atomic proposition.
Using a labeling map $\mathcal{L}: \Y\rightarrow \alphabeth$, we can define temporal logic specifications over the output of a system $\M$. Each output trace $\mathbf y \!=\! y_0,y_1,y_2,\ldots$ of $\M$ can be translated to a word via $\word \!=\! \mathcal{L}(\mathbf y)$. We say that a system $\M$ (e.g., $\M:=\agent\times\env$) satisfies the specification $\psi$ with probability of at least $p$ if $\P(\M\satisfies \psi):=\mathbb{E}[\mathcal{L}(\mathbf{y})\satisfies \psi] \ge p$, where $\mathbf{y}$ is any output trace emitted by $\M$.

\subsection{Simulated Environments}
Instead of performing exhaustive computations on the potentially high-dim\-ensional $\env$ --- in the spirit of SA --- we replace $\env$ with a lower-dimensional and less complex dynamics model, which we aptly call the \emph{simulator} $\simulator$ .
To compensate for the loss in dynamic resemblance, we invoke the concept of \emph{red teaming}. More specifically, we equip the simulator $\simulator$ with an adversary $\adv:\X^\simulator\times\O^\simulator\times\borel{\D^\simulator}\rightarrow[0,1]$ that perturbs $\simulator$ with adversarial disturbances $d^\simulator_t\sim\adv(\cdotx| x^\simulator_t,o^\simulator_t)$.
The resulting system is of the form
\begin{align}
    \simulator(\theta^\env,\adv):&\left\{\begin{array}{ll}
        x^\simulator_{t+1} &= f^\simulator(x^\simulator_t, o^\simulator_t; \theta^\env) + d^\simulator_t +  w^\simulator_t\\[.2em]
	y^\simulator_t &= h^\simulator(x^\simulator_t)
    \end{array}\right., && \begin{array}{ll}w^\simulator_t&\sim \mathcal{N}_{\W^\simulator}(\cdotx|0,\Sigma^\env)\\[.2em]
    d^\simulator_t&\sim\adv(\cdotx| x^\simulator_t,o^\simulator_t)\end{array}.\label{eq:simulator}
\end{align}
We will require that $x^\simulator$ retains all information relevant w.r.t. the output by assuming $h^\env=h^\simulator\circ P$ via some measurable map $P:\X^\env\rightarrow\X^\simulator$.
Without loss of generality, we assume that the disturbance $d^\simulator_t\sim\adv(\cdotx)$ is additive.
Based on an ambiguity set $\advAmbSet$ of adversaries $\adv\in\advAmbSet$, \eqref{eq:simulator} defines a family of simulator models $\smash{\simulator(\Theta,\advAmbSet)}:=\smash{\{\simulator(\theta^\env,\adv)\mid\theta^\env\in\Theta^\env,\adv\in\advAmbSet\}}$.
Analogous to $(\agent\times\env)(\theta)$, we define $\smash{(\agent\times_a\simulator)(\theta,\adv)}:=\smash{\agent(\theta^\agent)\times_a\simulator(\theta^\env,\adv)}$,
where $\times_a:(x^{\agent},x^\simulator)\mapsto(F(x^{\agent}),x^\simulator)=(o^\simulator,o^\agent)$ with some measurable map $F:\X^\agent\rightarrow\O^\simulator$.
The additional mapping $F$ will grant us the flexibility to apply MOR to $\agent$ (see Theorem~\ref{thm:mainthm}).

\begin{rexample}
Let us assume that for the scenario at hand the \emph{relevant} dynamics of $\env$ are the terms concerning its longitudinal movement. These could be, for example, scenarios such as the ones shown in Fig.~\ref{fig:situations}. Hence, we can substitute $\env$ by choosing a simulator of the form
\begin{align}
    \simulator(\theta^\env,\adv): \left\lbrace
    \begin{array}{ll}	
        \begin{bmatrix}
            v^\simulator_{t+1}\\
            s^\simulator_{t+1}
        \end{bmatrix}
        &= \begin{bmatrix}
            \theta^\env & 0\\
            0\footnotemark & 1
	\end{bmatrix}x_t^\simulator + \begin{bmatrix}
	    a_4&0&0\\a_5&0&0
	\end{bmatrix} o_t^\simulator + \adv(\cdotx) + w_t^\simulator\\[.8em]
    y_t^\simulator &= x_t^\simulator
    \end{array}
    \right.,\label{eq:simulator_car}
\end{align}\footnotetext{$\tau v_t^\simulator$ was absorbed into \eqref{eq:disturbance_model} as the disturbance would otherwise become multiplicative.}
with state $x^\simulator_t:=[v^\simulator_t,s^\simulator_{t}]\T\in\X^\simulator$,
observation $o_t^\simulator\in\X^\agent$,
process noise $w^\simulator_t\sim p^\env_{w,v} p^\env_{w,s_x}$, and adversarial disturbance $d_t^\simulator=\adv(x^\simulator_t,o^\simulator_t)$, elicited by a deterministic nonlinear adversary of the form
\begin{equation}
    \adv \in \advAmbSet:=\left\lbrace(x^\simulator_t,o^\simulator_t)\mapsto\begin{bmatrix}
        0\\
    \tau v^\simulator_t\cos(\beta+\Psi)\end{bmatrix} \Bigg\vert\, \beta\in\X_\beta^\env,\, \Psi\in\X_\Psi^\env\right\rbrace.\label{eq:disturbance_model}
\end{equation}
To reduce complexity further, we also choose to use a reduced-order model for the agent $\agent$ in \eqref{eq:agent_car}:
\begin{align}
    \agentred(\theta^\agent): \left\lbrace
    \begin{array}{ll}	
        \begin{bmatrix}
            v^{\agentred}_{t+1}\\
            s^{\agentred}_{t+1}
        \end{bmatrix}
        &= \begin{bmatrix}
            \theta^\agent&0\\
            \tau&1
	\end{bmatrix}x^{\agentred}_{t} + \begin{bmatrix}
	    \tau\\0
	\end{bmatrix}u^{\agentred}_{t} + w^{\agentred}_{t}\\[.8em]
    y^{\agentred}_{t} &= x^{\agentred}_{t}
    \end{array}
    \right.,\label{eq:agent_car_reduced}
\end{align}
with state \mbox{$x^{\agentred}_{t}:=[v^{\agentred}_{t},s^{\agentred}_{t}]\T\in\X^{\agentred}_{v}\times\X^{\agentred}_{s}$}, control input $u^{\agentred}_{t}\in\U^\agent$, and process noise $w^{\agentred}_{t}\sim p^\agent_{w,v}p^\agent_{w,s}$.
\end{rexample}

To obtain end-to-end guarantees, establishing a formal connection between the simulator $\simulator$ and the environment $\env$ is crucial. Whilst suitable notions are available for generic MDPs, our setting is more complex.
We build on the work \cite{schon2024bayesian}, which formulates stochastic simulation relations for a parametrized MDP $\M(\theta)$ by establishing a partial coupling between a set $\smash{\{\M(\theta)\mid\theta\in\Theta\}}$ and a nominal model $\M(\hat\theta)$.
Here, we intentionally extend $\M$ by an ambiguous adversary $\adv\in\advAmbSet$, i.e., $\M(\theta,\adv)$, as it allows us to simplify $\M$ itself. As a result, the nominal $\M(\hat\theta,\hat\adv)$ used for control synthesis can be much simpler.

\subsection{Problem Formulation}
Consider an agent $\agent$ in an environment $\env$.
Our goal is to design a controller $\C$ for $\agent$ such that the composed system $(\C\times\agent)\times\env$ satisfies a given specification $\psi$ with a probability not lower than some fixed $p\in(0,1)$.
To make the control design scalable, we leverage the notion of SA. 
In other words, although the concrete environment $\env$ may be a complex, high-dimensional system, we will assume that the dynamics \emph{relevant} to the interaction with the agent $\agent$ in a particular situation $i\in\mathcal{I}:=\{1,2,\ldots\}$ can be distilled in a much simpler form.
For instance, we can coarsely partition the domain $\X^\env$ of $\env$ to define finitely many situation-dependent $\env_i$ as restrictions of $\env$ to $\X^{\env_i}$ such that $\cup_{i\in\mathcal{I}} \X^{\env_i}=\X^\env$.
Then, for each partition $\X^{\env_i}$, we design a simulator model $\simulator_i$ alongside a set of adversaries $\advAmbSet$ such that $\simulator_i(\Theta,\advAmbSet)$ encapsulates a valid overapproximation of all the dynamical features of $\env_i$ when operating in $\X^{\env_i}$. We may therefore also call $\X^{\env_i}$ the \emph{region of validity} (RoV) for the simulator $\simulator_i$.
In fact, it might suffice for $\simulator_i$ not only to be much less complex and lower dimensional than $\env_i$, but modeling all stochastic features of $\env_i$ precisely might not be necessary either.
We will call such an environment $\env_i$ that associates a simpler simulator $\simulator_i$ on RoV $\X^{\env_i}$ \emph{elemental}.
In this work, we focus on a singular elemental environment $\env$ and
establish a probabilistic relation that allows us to compute a robust controller for $\agent\times\env$ via $\agent\times_a\simulator$ as a proxy.
\begin{prob}
    \label{prob:problem1}
    Let a simulator $\simulator$ for an elemental environment $\env$ with RoV $\X^\env$ be given such that $\simulator$ exhibits a richer behavior than $\env$.
    For a global scLTL specification $\psi$ and a corresponding threshold $p\in(0,1)$, design a controller $\C$ via $\agent\times_a\simulator$ such that 
    $\P((\C\times\agent)\times\env\satisfies\psi) \geq p.$
\end{prob}

\section{System Interrelations}\label{sec:relations}
In this work, we use two different types of relations between dynamical systems, \emph{behavioral inclusions} (BIs) and \emph{sub-simulation relations} (SSRs), that are introduced in Subsections~\ref{sec:behavioral_inclusion} and \ref{sec:stoch_simulation_relation}, respectively.
In order to formalize these notions, 
we model the agent, environment, and simulator \eqref{eq:agent}-\eqref{eq:simulator} as \emph{general Markov decision processes} (gMDPs). 
\begin{definition}[General Markov Decision Process (gMDP)]\label{def:gMDP}
A gMDP is a tuple $(\X,\X_0,\U,\O,\D,\Tr,\Y,h)$, comprising
$\X$, a measurable state space of states $x\in\X$;
$\X_0\subset\X$, a set of initial states $x_0\in\X_0$;
$\U$, a measurable input space of inputs $u\in\U$;
$\O$, a measurable set of observations $o\in\O$;
$\D$, a measurable set of disturbances $d\in\D$;
$\Tr:\X\times\U\times\O\times\D\times\mathcal B(\X)\rightarrow[0,1]$, a probability kernel; 
$\Y$, a measurable output space of outputs $y\in\Y$, decorated with a metric $\mathbf d_\Y$; and 
$h:\X\rightarrow\Y$, a measurable output map.
\end{definition}
We will overload the notation and omit tuple elements whenever they are nil.
For instance, in the remainder of the manuscript, we will treat the agent and environment as gMDPs of the form $\agent=(\X^\agent,\X_0^\agent,\U^\agent,\O^\agent,\Tr^\agent,\Y^\agent,h^\agent)$ and $\env=(\X^\env,\X_0^\env,\O^\env,\Tr^\env,\Y^\env,h^\env)$, respectively.
As we focus on autonomous environments, we omit the input set $\U^\env:=\emptyset$. Furthermore, we assume that both $\agent$ and $\env$ do not receive any additional disturbances, i.e., $\D^\agent=\D^\env=\emptyset$.
The transition kernels $\Tr^\agent,\Tr^\env$ are obtained from \eqref{eq:agent}-\eqref{eq:environment}, respectively, via
\begin{equation*}
    \Tr^\M(d\xp^\M| z^\M;\theta^\M) := \int_{\W^\M} \delta_{f^\M(z^\M;\theta^\M)+w^\M}(\xp^\M)\; 
    \mathcal{N}_{\W^\M}(dw^\M|0,\Sigma^\M),
\end{equation*}
where $z^\M:=(x^\M,u^\M,o^\M)$ and $\M\in\{\agent,\env\}$.
Note that the resulting (para\-metrized) transition kernels are conditioned on uncertain parameters $\theta^\M\in\Theta^\M$.
The simulator \eqref{eq:simulator} is captured as a gMDP $\simulator=(\X^\simulator,\X^\simulator_0,\O^\simulator,\D^\simulator,\Tr^\simulator,\Y^\simulator,h^\simulator)$, receiving additional disturbances $d^\simulator\in\D^\simulator$ from an ambiguous adversary $\adv\in\advAmbSet$.

\subsection{Behavioral Inclusion}\label{sec:behavioral_inclusion}
We now formalize the relation that we will use to relate environments and simulators, utilizing the ambiguous adversary $\adv\in\advAmbSet$.

\begin{definition}[Behavioral Inclusion (BI)]\label{def:BI}
    Consider autonomous gMDPs $\M=(\X,\X_0,\O,\Tr,\Y,h)$, \mbox{$\Mh=(\Xh,\Xh_0,\Oh,\Dh,\Trh,\Y,\hat h)$}, an ambiguity set $\hat\advAmbSet$ of adversaries $\hat\adv:\Xh\times\Oh\times\borel{\Dh}\rightarrow[0,1]$, and measurable maps $P:\X\rightarrow\Xh$ and $F:\O\rightarrow\Oh$. For the relation $\rel:=\{(\hat x,x)\in\Xh\times\X\mid \hat x = P(x)\}$, if
    \begin{itemize}
	\item[(a)] $\Xh_0=P(\X_0)$; 
        \item[(b)] $\forall (\hat x,{x})\in\rel,\,\forall o\in\O,\,\exists\hat\adv\in\hat\advAmbSet,$ $\forall \hat Q\in\borel{\Xh}:\; \Trh(\hat Q| \hat x,F(o),\allowdisplaybreaks\hat\adv(\hat d| \hat x,F(o)))=\allowdisplaybreaks \Tr(Q|x,o)$, with $Q:=P^{-1}(\hat Q)$; and
        \item[(c)] $\forall (\hat x,x)\in\rel:\, h(x)=\hat h(\hat x)$;
    \end{itemize}
    then, we say that \emph{the behavior of $\M$ is included in $\Mh$}, denoted as \mbox{$\simreltwo{\Mh(\hat\advAmbSet)}{\M}{}{}$}.
\end{definition}
Intuitively, the behavior of a system $\M$ is included in $\Mh$ if the following conditions are met:
(a) All initial conditions of $\M$ are matched by $\Mh$.
(b) For all state pairs in the relation and all extrinsic observations $o\in\O$, there exists an adversary $\hat\adv\in\hat\advAmbSet$ resolving the nondeterminism of $\Mh$ such that the resulting probability measure $\Trh(\cdotx| \hat x,F(o),\hat\adv(\hat d| \hat x,F(o)))$ assigns a probability to an event $\hat Q$ that matches the probability of all equivalent events $Q:=P^{-1}(\hat Q)$ of $\M$, as quantified by $\Tr(\cdotx|x,o)$.
In other words, $Q:=P^{-1}(\hat Q)$ represents an \emph{equivalence class} of rich-in-detail events that are all mapped to the same abstract event $\hat Q$ via the map $P$.
Finally, (c) whilst in the relation, both systems emit equivalent outputs.

\medskip

At a first glance, the BI bears conceptual resemblance to the relations used in \cite[Definition~4.1]{Tabuada2009Verif} and for perception in \cite{astorga2023perception}. However, their formulations are substantially different as they are restricted to deterministic systems and do not specify how the relation can be established for systems of different dimensionality.
More similarly, our condition (b) can be seen as a generalization of the bisimulation for nondeterministic labeled Markov processes in \cite{budde2012theory} to antisymmetric relations $\rel$.
In fact, the map $P$ in the BI acts as a \emph{zigzag morphism}~\cite{desharnais2002bisimulation,edalat1999semi}, preserving transition probabilities across the two labeled Markov processes $\smash{\M,\Mh}$.
We refer to the early \cite{desharnais2003approximating} and \cite[Appendix~C]{haesaert2017verification}, which are situated in a similar relational context.
In effect, the BI imposes the strict ordering of $\M$ and $\smash{\Mh}$ according to their level of abstraction, as outlined in the intuition above.

\begin{rexample}
It is easy to verify that for the chosen RoV $\X^\env$ of $\env$ --- in particular, tight bounds on the slip angle $\beta^\env\in\X_\beta^\env$ and yaw angle $\Psi^\env\in\X_\Psi^\env$ --- the lateral dynamics become neglectable and the simulator $\smash{\simulator(\theta^\env,\adv)}$ in \eqref{eq:simulator_car} in conjunction with the ambiguous adversary $\adv\in\advAmbSet$ in \eqref{eq:disturbance_model} subsumes all the behavior of the environment $\env(\theta^\env)$ in \eqref{eq:env_car}, i.e., we have \mbox{$\simreltwo{\simulator(\theta^\env,\advAmbSet)}{\env(\theta^\env)}{}{}$} via the map $P:[0_{2\times3},I_2,0_{2\times1}]x^\env\mapsto x^\simulator$.
Note, that in contrast to the six-dimensional nonlinear $\env$ the simulator $\simulator$ is only two-dimensional, linear, and lifts the need for accurate knowledge of the underlying stochastic laws on $\beta^\env,\,\mathrm{d}\Psi^\env,\,\Psi^\env$, and $s_y^\env$.
In practice, a basic model such as \eqref{eq:simulator_car} is often supplied by the SA component of an autonomous agent~\cite{casablanca2024symaware}. Reachability techniques can be used to construct the corresponding adversary.
\end{rexample}

\begin{remark}[Non-probabilistic simulators]
    The adversary in \eqref{eq:disturbance_model} can be augmented to absorb the noise $w^\simulator_t\in\smash{\W^\env_{w,v}\times\W^\env_{w,s_x}}$ also, lifting the need for \emph{any} information on the stochastic laws of $\env$ and yielding a fully non-probabilistic simulator $\simulator$. 
    This would further trade scalability for conservatism.
\end{remark}

\subsection{Stochastic Simulation Relation}\label{sec:stoch_simulation_relation}
Before providing the definition of an SSR, we recall the instrumental definition of a sub-probability coupling between two probability measures \cite[Definition~5]{schon2024bayesian}.
\begin{definition}[Sub-probability coupling]\label{def:coupling}
    Given $\hat p\in \mathcal P(\Xh)$ and $p\in\mathcal P(\X)$,  $\rel\subset\borel{\Xh\times\X}$, and a value $\delta\in[0,1]$, we say that a sub-probability measure $v$ over $(\Xh\times\X , \borel{\Xh\times\X})$ with $v(\Xh\times\X)\geq 1-\delta$ is a \emph{sub-probability coupling} of $\hat p$ and $p$ over $\rel$ if
    \begin{itemize}
	\item[(a)] $v(\Xh\times\X)=v(\rel)$, i.e., $v$'s probability mass is located on $\rel$;
	\item[(b)] $\forall A\in\borel{\X}$:\, $v(\Xh\times A)\leq p(A)$; and
	\item[(c)] $\forall A\in\borel{\Xh}$:\, $v(A\times \X)\leq\hat p(A)$.
    \end{itemize}
\end{definition}

With this, we extend the original SSR Definition by \cite{schon2024bayesian} to accommodate a nondeterministic adversary.
\begin{definition}[($\varepsilon,\delta$)-Sub-Simulation Relation (SSR)]\label{def:ssr}
    Consider two gMDPs $\Mh=(\Xh,\Xh_0,\Uh,\Dh,\Trh,\hat h,\Y)$ and $\M=(\X,\X_0,\U,\D,\Tr,h,\Y)$, 
    a set $\hat\advAmbSet$ of adversaries $\hat\adv:\Xh\times\borel{\Dh}\rightarrow[0,1]$, an adversary $\adv:\X\times\borel{\D}\rightarrow[0,1]$,
    a measurable relation $\rel\subset\borel{\Xh\times\X}$, and an interface function $\InFu: \Xh\times\X\times\Uh\times\borel{\U}\rightarrow[0,1]$. If there exists a sub-prob. kernel $\Wt:\Xh\times\X\times\Uh\times\Dh\times\D\times\borel{\Xh\times\X}\rightarrow[0,1]$ s.t.
    \begin{itemize}
	\item[(a)] $\forall x_0\in\X_0,\,\exists\hat{x}_0\in\Xh_0:\,(\hat x_{0},{x}_{0})\in\rel$;
	\item[(b)] $\forall(\hat x, x)\in\rel$, $\forall\hat u\in\Uh$, there exists an $\hat \adv\in\hat\advAmbSet$ such that $\Wt(d\xhp\times d\xp|\hat x,x,\allowbreak\hat u,\allowbreak\hat\adv(\hat d|\hat x),\allowbreak\adv(d|x))$ is a sub-probability coupling of $\Trh(d\xhp|\hat x,\hat u,\hat\adv(\hat d|\hat x))$ and $\Tr(d\xp|\allowbreak x,\allowbreak\InFu(u|\hat x,x,\hat u),\allowbreak\adv(d|x))$ over $\rel$ with respect to $\delta$; and
	\item[(c)] $\forall(\hat x,x)\in\rel:\,\mathbf{d}_\Y(\hat h(\hat x),h( x))\leq\varepsilon$;
    \end{itemize}
    then, $\M$ is in an SSR with $\Mh$, denoted as $\simrel{\Mh(\hat\advAmbSet)}{\M(\adv)}{\varepsilon}{\delta}$.
\end{definition}
This extended SSR definition recovers the previous \cite[Definition~6]{schon2024bayesian} for the undisturbed case, where $\smash{\hat\advAmbSet}=\{0\}$ and $\adv=0$. In the following, with a slight abuse of notation, we may write $\smash{\simrel{\Mh(\hat\adv)}{\M(\adv)}{\varepsilon}{\delta}}$ to indicate that $\smash{\hat\advAmbSet}=\{\hat\adv\}$.

\smallskip

The intuition behind the three conditions of Definition~\ref{def:ssr} on a composed system $\Mh\times\M$ evolving on the product space $\Xh\times\X$ is as follows. Note that $\rel$ defines a subset of $\Xh\times\X$. According to condition (a), both systems start in $\rel$ upon initialization. In any subsequent time step, once in $\rel$, condition (b) certifies that the systems remain in $\rel$ for the next time step with a probability of at least $(1-\delta)$ for all control inputs $\hat u\in\Uh$ and adversaries $\hat\adv\in\hat\advAmbSet$. Finally, given the two systems are in $\rel$, the corresponding outputs $\smash{\hat y}:=\hat h(\hat x)$ and $ y:= h( x)$ are $\varepsilon$-close (condition (c)).

\section{Assume-Guarantee Contracts}\label{sec:agc}
In this section, we associate the previously introduced systems and system interrelations within the framework of \emph{assume-guarantee contracts} (AGCs). To this end, \emph{assumptions} and \emph{guarantees} resemble qualitative types of dynamical systems --- conceptually similar to \cite{Shali2022AGC}.
\begin{definition}[Assumption]
    Given gMDPs $\M$ and $\Mh$, $\Mh(\advAmbSet)$ is an \emph{assumption} of $\M$ (on the RoV $\X^\M$) if $\simreltwo{\Mh(\advAmbSet)}{\M}{}{}$.
\end{definition}
In essence, an assumption of a system $\M$ is a model $\Mh$ that subsumes all its behavior (on a given RoV).
Thus, a simulator $\simulator$ is as an assumption of a complex environment $\env$ if $\simreltwo{\simulator(\advAmbSet)}{\env}{}{}$
 --- \emph{qualifying} $\simulator$ as a formally valid surrogate.

\smallskip

We proceed with the definition of \emph{$(\varepsilon,\delta)$-guarantees}, which, conversely, are specified for composed systems,
and render the systems qualified \emph{proxies}.
\begin{definition}[$(\varepsilon,\delta)$-Guarantee]
    Let $\varepsilon\geq0$ and $\delta\in[0,1]$.
    Given gMDPs $\M_1$ and $\M_2$, a \emph{$(\varepsilon,\delta)$-guarantee of $\M_1$ in $\M_2$} is a gMDP $\gua$ such that
    $\simrel{\M_{1\!}\times\M_{2\!}}{\gua}{\varepsilon}{\delta}.$
\end{definition}

We are ready to present the main result of this paper: We cast a formal AGC that 
\emph{assumes} that the simulator $\smash{\simulator(\advAmbSet)}$ is a valid surrogate for $\env$ (on the RoV $\X^\env$) and 
renders a reduced-order system $\smash{\agentred\times\simulator}$ a proxy of $\agent\times\env$ --- the \emph{guarantee}.
For a given relation $\rel\subset\borel{(\X^{\agentred}\!\times\!\X^\simulator)\!\times\!(\X^{\agentred}\!\times\!\X^\simulator)}$, we define composite relations on $\borel{(\X^{\agentred}\!\times\!\X^\simulator)\!\times\!(\X^{\agent}\!\times\!\X^\simulator)}$ and $\borel{(\X^{\agentred}\!\times\!\X^\simulator)\!\times\!(\X^{\agent}\!\times\!\X^\env)}$ , i.e.,
\begin{align*}
   \rel_a&:=\rel\circ\times_a,&&\times_a:(x^{\agent},x^\simulator)\mapsto(F(x^{\agent}),x^\simulator) &&\text{and}
   \\
    \rel_b&:=\rel\circ\times_b,&&\times_b:(x^{\agent},x^\env)\mapsto(F(x^{\agent}),P(x^\env)),&&
\end{align*}
respectively, with maps $F:\borel{\X^\agent}\rightarrow\X^{\agentred}$ and $P:\borel{\X^{\env}}\rightarrow\X^\simulator$.
\begin{theorem}[Proxy]\label{thm:mainthm}
    Let gMDPs $\agent$, $\agentred$, and $\env$ be given. Let $\simulator(\advAmbSet)$ be an assumption of $\env$, i.e., $\simreltwo{\simulator(\advAmbSet)}{\env}{}{}$ via $P$.
    Then, for $\varepsilon\geq0$ and $\delta\in[0,1]$:
    \begin{subequations}
    \begin{align}
    	&
     \forall\adv\in\advAmbSet:\,
        \simrel{(\agentred\times\simulator)(\adv))}{(\agent\times_a\simulator)(\adv)}{{\varepsilon}}{\delta}{}{}\label{eq:inductionmain_precond}
     \\
    	&\hspace{84pt}
        \implies\,\, 
        \simrel{(\agentred\times\simulator)(\advAmbSet)}{\agent\times\env}{\varepsilon}{\delta},
     \label{eq:inductionmain_postcond}
    \end{align}\label{eq:inductionmain}%
    \end{subequations}
    with relations $\rel_a$ and $\rel_b$, respectively. 
\end{theorem}
Theorem~\ref{thm:mainthm} allows us to reason about $\agent\times\env$ via $\agentred\times\simulator$ as a proxy without requiring us to compute the error bounds $\varepsilon,\delta$ considering the potentially high-dimensional domain of $\agent\times\env$. Instead, we compute the error bounds for the relation between $\agent\times_a\simulator$ and $\agentred\times\simulator$ --- both involving the simpler simulator $\simulator$.
\begin{remark}[Reduced conservatism]
    Consider $\simrel{\agentred}{\agent}{\varepsilon_1}{\delta_1}{}{}$ and $\simrel{\simulator}{\env}{\varepsilon_2}{\delta_2}{}{}$ for some $\varepsilon_1,\varepsilon_2\geq0$ and $\delta_1,\delta_2\in[0,1]$, which also implies \eqref{eq:inductionmain_postcond} with $\varepsilon=\varepsilon_1+\varepsilon_2$ and $\delta=1-(1-\delta_1)(1-\delta_2)$ \cite[Theorem~1]{Schoen2023GMM}. 
    In comparison, our result is less restrictive, since they assume the worst-case behavior of the other system to establish local simulation relations. 
\end{remark}
Since for each $\adv\in\advAmbSet$ the simulator $\simulator(\adv)$ in Theorem~\ref{thm:mainthm} is a (potentially deterministic) low-dimensional representation of the environment $\env$, establishing the antecedent \eqref{eq:inductionmain_precond} is much easier than the consequent \eqref{eq:inductionmain_postcond}.
Theorem~\ref{thm:mainthm} then provides the guarantees that the closed-loop agent system will behave reliably in the actual environment without ever having to compute their interaction directly. 
In contrast to traditional compositional results  relying on assumptions of dissipativity, low gain, or contractivity, we only require the assumption $\simreltwo{\simulator(\advAmbSet)}{\env}{}{}$ to hold.
Thus, our AGC can allow for computationally tractable control synthesis. 

\begin{remark}[Networks of subsystems]
Most work on compositionality is focused on networks of $N>>2$ subsystems, where the goal is to compute local controllers for each subsystem \cite{nuzzo2019stochastic,ghasemi2020compositional,lavaei2022compositional,Schoen2023GMM,mallik2018compositional,kazemi2024assume}.
We present our AGC result for the two subsystems $\agent$ and $\env$ whilst remarking that it can be easily applied to $\agent$ in a network of autonomous subsystems, by defining the rest of the network as its environment. This is the case, for example, when designing a controller for an agent interacting with other autonomous entities.
For the more general case of multi-agent control, prior work computes local controllers on the decoupled subsystems of dimensionality $d$ assuming weak interconnection, resulting in a complexity of $\mathcal{O}(Nn^{2d})$, where $d$ is the maximum dimension of the subsystems.
We can apply the same setup as before whilst adjusting local controllers iteratively, which scales with $\mathcal{O}(N n^{2(d+d')})$, where $d'$ is the maximum dimension of the reduced-order model of the environment for each active subsystem.
\end{remark}

\section{Bridging the Gap}\label{sec:simtoreal}
In this section, we provide an abstraction-based solution to Problem~\ref{prob:problem1}.
For this, we leverage the AGC result in Theorem~\ref{thm:mainthm} to establish a formal link between the original system $(\agent\times\env)(\theta)$ and a \emph{finite} nominal model $(\agentred\times\simulator)_f(\hat\theta,\hat\adv)$.
We synthesize a controller via $(\agentred\times\simulator)_f(\hat\theta,\hat\adv)$ that can be readily applied to $\agent\times\env$, by robustifying it against parametric uncertainty, adversarial disturbance, and the discretization error.

\subsection{Situational Awareness}\label{sec:situational_awareness}
Instead of performing computations on the actual system $(\agent\times\env)(\theta)$,
we replace it with the less complex simulated system $(\agentred\times\simulator)(\theta,\advAmbSet)$.
We assume that the simulator $\simulator$ is constructed such that $\simreltwo{\simulator(\advAmbSet)}{\env}{}{}$.
Theorem~\ref{thm:mainthm} establishes a formal link between the two systems of the form $\simrel{({\agentred}\times\simulator)(\theta,\advAmbSet)}{({\agent}\times\env)(\theta)}{\varepsilon}{\delta}$ if the antecedent $\simrel{(\agentred\times\simulator)(\theta,\adv)}{(\agent\times_a\simulator)(\theta,\adv)}{{\varepsilon}}{\delta}{}{}$ holds for all $(\theta,\adv)\in\Theta\times\advAmbSet$.
To establish the antecedent,
we can use a \emph{coupling compensator} for model-order reduction (MOR).
We propose the following partial compensator.

\begin{proposition}[Partial MOR coupling compensator]\label{prop:MOR_compensator_partial}
    Consider two gMDPs $\M(\theta,\adv)=(\X,\X_0,\U,\D,\Tr,h,\Y)$ and $\M_r(\theta,\adv_r) = (\X_r, \X_{r,0}, \U_r, \D_r, \Tr_r, h_r, \Y)$
    with a mapping $F:\X\rightarrow\X_r$ such that $x_r = F(x)$, an interface function $u_t\sim\InFu(\cdotx| x_t, x_{r,t}, u_{r,t})$, and a relation
    $\rel_r:= \left\lbrace(x_r,x) \in \X_r\times \X \,\left\vert\, \norm{F(x)-x_r}_{D_r}\right.\leq\varepsilon_r\right\rbrace,$
    where $\norm{z}_{D_r}:=\smash{\sqrt{z\T D_r z}}$ denotes the two norm with weight matrix $D_r$.
    If there exist $\varepsilon_r\geq0$ and $\delta_r:\X\times\U_r\rightarrow[0,1]$ such that
    \begin{itemize}
        \item $\varepsilon_r\geq\sup_{(x_r,x)\in\rel_r}\norm{h(x)-h_r(x_r)}$; and
        \item there exists a sub-prob. kernel $\Wt:\X_r\times\X\times\U_r\times\borel{\X_r\times\X}\rightarrow[0,1]$, such that $\forall(x_r,x)\in\rel_r$ and $\forall u_r\in\U_r$ the resulting measure $\Wt(\new{dx_{r+}\times d\xp}|x_r,x,u_r)$ is a sub-prob. coupling of $\Tr_r(\new{dx_{r+}}|x_r,u_r,\adv_r(d_r|\cdotx);\theta)$ and $\Tr(\new{d\xp}|\allowbreak x,\allowbreak\InFu(u|\cdotx),\allowbreak\adv(d|\cdotx);\theta)$ over $\rel_r$ with $\Wt(\rel_r)\geq1-\delta_r(x,u_r)$;
    \end{itemize}
    then, $\simrel{\M_r(\theta,\adv_r)}{\M(\theta,\adv)}{\varepsilon_r}{\delta_r}$.
\end{proposition}
The above result provides conditions for the computation of the error parameters $\varepsilon_r$ and $\delta_r$ of the simulation relation between $\M$ and $\M_r$. We give an examination of its distinction to prior works in 
\iflong
Appendix~\ref{app:MOR_coupling_compensator}.
\else
\cite[Appendix~B]{extendedVersion}.
\fi
Note that the resulting $\delta_r$ is dependent on the original state $x$. This dependence can be resolved by taking the worst case w.r.t. $x\in\X$ (if feasible) or --- given the relevant variables from $x$ are observed upon runtime --- adding (a subset of) $x$ as an input to $\M_r$.

\begin{rexample}
    Based on Proposition~\ref{prop:MOR_compensator_partial}, we establish $\simrel{(\agentred\times\simulator)(\theta,\adv)}{(\agent\allowbreak\times_a\simulator)(\theta,\adv)}{{\varepsilon_1}}{\delta_1}{}{}$, $(\theta,\adv)\in\Theta\times\advAmbSet$, 
    for parameters $a_1=0.2$, $b=0.001$, and $a_2=a_3=a_4=a_5=0$,
    with $\varepsilon_1=0$ and $\delta_1=0\%$.
    See 
    \iflong
    Appendix~\ref{app:reduced_agent} 
    \else
    \cite[Appendix~C.3]{extendedVersion}
    \fi
    for more details.
    We summon Theorem~\ref{thm:mainthm} to infer the relation $\simrel{({\agentred}\times\simulator)(\theta,\advAmbSet)}{({\agent}\times\env)(\theta)}{\varepsilon_1}{\delta_1}$.
\end{rexample}

\subsection{Ambiguity Compensation}
The system $(\agentred\times\simulator)(\theta,\adv)$ incorporates two types of ambiguity, arising from the parametric uncertainty $\theta\in\Theta$ and the adversary $\adv\in\advAmbSet$.
In fact, our aim is to relate a nominal system $\smash{(\agentred\times\simulator)(\hat\theta,\hat\adv)}:=(\Xh,\Xh_0,\Uh,\Dh,\hat\adv,\Trh,\Yh,\hat h)$ with a fixed parameter estimate $\smash{\hat\theta}$ and a (potentially simplified) nominal adversary $\hat\adv:\smash{\Xh}\times\smash{\borel{\Dh}}\rightarrow[0,1]$ to the ambiguity set of systems $(\agentred\times\simulator)(\Theta,\advAmbSet)$.
To this end, we start by rewriting the dynamics of $(\agentred\times\simulator)(\theta,\adv)$ as follows:
\begin{align*}
    x_{t+1} &=
    \underbrace{\hat f(\hat x_t,\hat u_t;\hat\theta)+\hat \adv(\cdotx|\hat x_t)}_{\text{nominal dynamics}}
    + \underbrace{\Delta(\cdotx) + w_t}_{\text{disturbance}}, && w_t\sim p_w(\cdotx),\\
    &= \hat f(\hat x_t,\hat u_t;\hat\theta)+\hat\adv(\cdotx|\hat x_t) + \hat w_t, && \hat w_t:= w_t + \Delta(\cdotx),
\end{align*}
where the noise distribution $p_w$ is offset by the error dynamics
\begin{equation}
    \Delta(\cdotx):=(f(x,u;\theta)+\adv(\cdotx|x))-(\hat f(\hat x,\hat u;\hat\theta)+\hat\adv(\cdotx|\hat x)).\label{eq:Delta}
\end{equation} 
Based on the work by \cite{schon2024bayesian},
we establish the following extended result to compensate both the parametric and adversarial ambiguity.
\begin{theorem}[Ambiguity compensation]\label{thm:ambiguity_compensation}
    For the models $(\agentred\times\simulator)(\theta,\adv)$ and $(\agentred\times\simulator)(\hat\theta,\hat\adv)$ with $(\theta,\adv)\in\Theta\times\advAmbSet$, we have
    that $\simrel{(\agentred\times\simulator)(\hat\theta,\hat\adv)}{(\agentred\times\simulator)(\theta,\adv)}{\varepsilon}{\delta}$ with interface function $u_t=\hat{u}_t$, relation
    $\rel:=\{(\hat x,x)\in\Xh\times\X\mid\hat x=x\}$,
    state mapping
    $\xhp = \xp$,
    parameter $\varepsilon = 0$, and with the offset $\Delta$ defined in \eqref{eq:Delta}
    \begin{equation}
        \delta(\hat x,\hat u;\hat\theta,\hat\adv) = 1-2\,\cdf{-\frac{1}{2}
        \sup_{\theta\in\Theta,\adv\in\advAmbSet}
        \norm{\Delta(\hat x,\hat u;\hat\theta,\theta,\hat\adv,\adv)}}.\label{eq:delta}
    \end{equation}
\end{theorem}

\begin{remark}[Disturbance refinement]
    The nominal disturbance model $\hat\adv$ can reduce the conservativeness of the approach by acting as a virtual disturbance feedback.
    To make the approach even less conservative, the actual disturbance based on $x^\env$ can be inferred and can be fed back into the nominal model via an additional input, as stated below Proposition~\ref{prop:MOR_compensator_partial}.
\end{remark}

\begin{rexample}
    We select a nominal model $(\agentred\times\simulator)(\hat\theta,\hat\adv)$ by fixing $\hat\theta:=\smash{(\hat{\theta}^\agent,\hat{\theta}^\env)}=(0.8, 0.8)$ and a linearized nominal adversary $\hat\adv:(x^{\simulator}_t,o^{\simulator}_t)\mapsto[0,\tau v^{\simulator}_t]\T.$
    Using Theorem~\ref{thm:ambiguity_compensation}, we obtain $\smash{\simrel{(\agentred\times\simulator)(\hat\theta,\hat\adv)}{(\agentred\times\simulator)(\theta,\adv)}{\varepsilon_2}{\delta_2}}$, with $\varepsilon_2=0$ and $\delta_2$ as a function of the initial state of $(\agentred\times\simulator)(\smash{\hat\theta},\hat\adv)$ with a maximum of $3.64\%$. See
    \iflong
    Appendix~\ref{app:ambiguity_compensation} 
    \else
    \cite[Appendix~C.4]{extendedVersion}
    \fi
    for further details.
\end{rexample}

\subsection{Embedding with Situational Awareness}
Given an SA block discriminating between situations $\{\env_i\}_{i\in\mathcal{I}}$ probabilistically, we may want to combine guarantees from multiple simulators $\{\simulator_i\}_{i\in\mathcal{I}}$ (see Fig~\ref{fig:idea}). 
\begin{theorem}[Situational awareness]
    Let $P_1,P_2,\ldots$ be the probabilities of the agent $\agent$ operating in the situations $\env_1,\env_2,\ldots$, and let $\{\simulator_i(\advAmbSet_i)\}_{i\in\mathcal{I}}$ be assumptions of $\{\env_i\}_{i\in\mathcal{I}}$. 
    Let $p_i$ and $\C_i$, respectively, be the robust satisfaction probability and robust controller of $\agent\times\env_i$ obtained via Theorem~\ref{thm:mainthm}. Then, we have for all $i\in\mathcal{I}$ that
    $\P((\C_i\times\agent)\times\env_i\satisfies\psi) \geq p^\ast$, where $p^\ast:=\sum_{i\in\mathcal{I}}P_i p_i$.
\end{theorem}

\section{Case Study}\label{sec:experiments}
\begin{SCfigure}
    \def\svgwidth{.5\linewidth}
\begingroup%
  \makeatletter%
  \providecommand\color[2][]{%
    \errmessage{(Inkscape) Color is used for the text in Inkscape, but the package 'color.sty' is not loaded}%
    \renewcommand\color[2][]{}%
  }%
  \providecommand\transparent[1]{%
    \errmessage{(Inkscape) Transparency is used (non-zero) for the text in Inkscape, but the package 'transparent.sty' is not loaded}%
    \renewcommand\transparent[1]{}%
  }%
  \providecommand\rotatebox[2]{#2}%
  \newcommand*\fsize{\dimexpr\f@size pt\relax}%
  \newcommand*\lineheight[1]{\fontsize{\fsize}{#1\fsize}\selectfont}%
  \ifx\svgwidth\undefined%
    \setlength{\unitlength}{686.83634949bp}%
    \ifx\svgscale\undefined%
      \relax%
    \else%
      \setlength{\unitlength}{\unitlength * \real{\svgscale}}%
    \fi%
  \else%
    \setlength{\unitlength}{\svgwidth}%
  \fi%
  \global\let\svgwidth\undefined%
  \global\let\svgscale\undefined%
  \makeatother%
  \begin{picture}(1,0.45711565)%
    \lineheight{1}%
    \setlength\tabcolsep{0pt}%
    \put(0,0){\includegraphics[width=\unitlength,page=1]{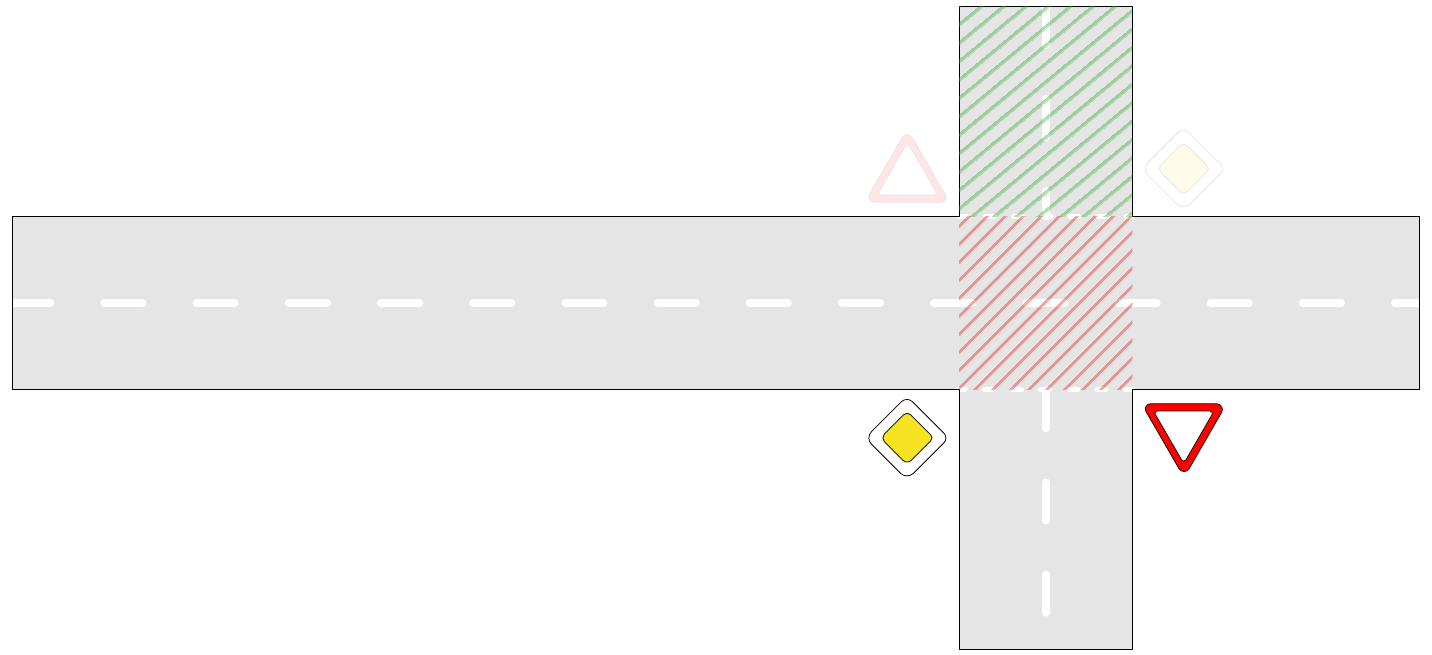}}%
    \put(0.73258965,0.2388997){\color[rgb]{0.8627451,0,0}\makebox(0,0)[t]{\lineheight{1.25}\smash{\begin{tabular}[t]{c}$P_C$\end{tabular}}}}%
    \put(0.73258965,0.37136764){\color[rgb]{0,0.61176471,0.00784314}\makebox(0,0)[t]{\lineheight{1.25}\smash{\begin{tabular}[t]{c}$P_T$\end{tabular}}}}%
    \put(0,0){\includegraphics[width=\unitlength,page=2]{Minimial_example.pdf}}%
    \put(0.20952308,0.14833147){\color[rgb]{0,0,0}\makebox(0,0)[t]{\lineheight{1.25}\smash{\begin{tabular}[t]{c}$\mathbf{E}$\end{tabular}}}}%
    \put(0.81533172,0.07295429){\color[rgb]{0,0,0}\makebox(0,0)[t]{\lineheight{1.25}\smash{\begin{tabular}[t]{c}$\mathbf{A}$\end{tabular}}}}%
  \end{picture}%
\endgroup%

    \caption{Intersection alongside the regions associated with collision $P_{C}$ and target $P_{T}$.}
    \label{fig:intersection}
\end{SCfigure}

Consider the scenario shown in Fig.~\ref{fig:intersection}, involving two vehicles --- $\agent$ and $\env$ --- at a traffic intersection. The corresponding dynamics are given in \eqref{eq:agent_car}-\eqref{eq:env_car}.
Our goal is to design a controller for the agent vehicle $\agent$, which wants to pass the intersection to continue its journey. It has to yield to the environment vehicle $\env$, which is traveling on the priority road. Based on the behavior of $\env$, the agent may have time to pass the intersection \emph{before} $\env$, or it may go \emph{after} $\env$ to avoid a collision.
In scLTL, this specification is written as $\psi:= (P_S \andltl\notltl P_C) \Until P_T$, where $P_S$ is a safety condition on the bounded domain $\X^\agent\times\X^\env$.
Note, that $\psi$ can \emph{not} be decomposed into local sub-specifications for $\agent$ and $\env$, rendering most traditional compositional approaches inapplicable.

\begin{figure}
    \centering
    \begin{subfigure}[b]{.49\linewidth}
    \includegraphics[width=\linewidth]{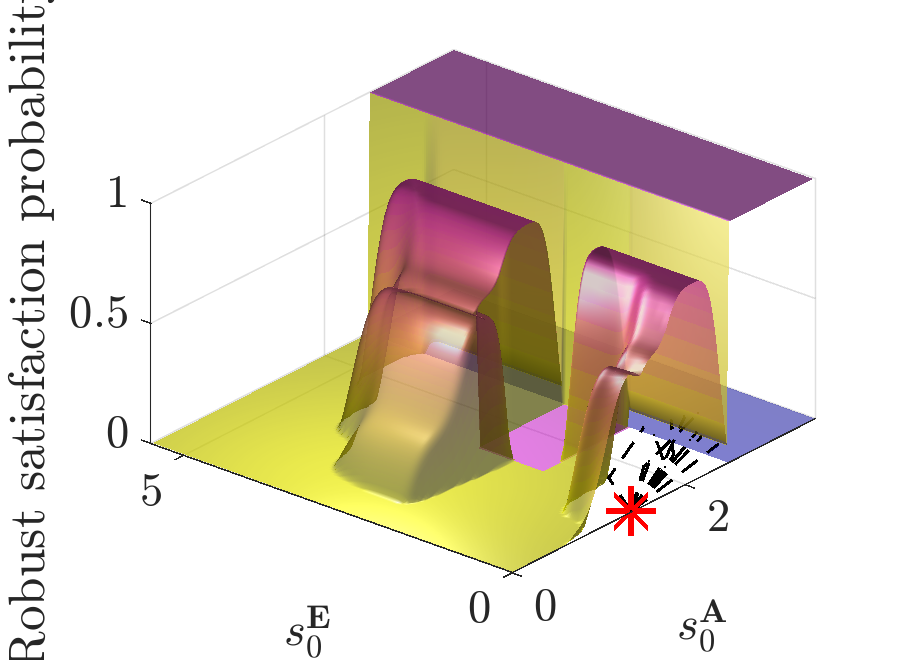}
    \caption{\mbox{$(v_0^\agent,v_0^\env)=(2.960,2.296)$} and $(s_0^\agent, s_0^\env)=\allowbreak(1.387,0.014)$}
    \end{subfigure}
    \hfill
    \begin{subfigure}[b]{.49\linewidth}
    \includegraphics[width=\linewidth]{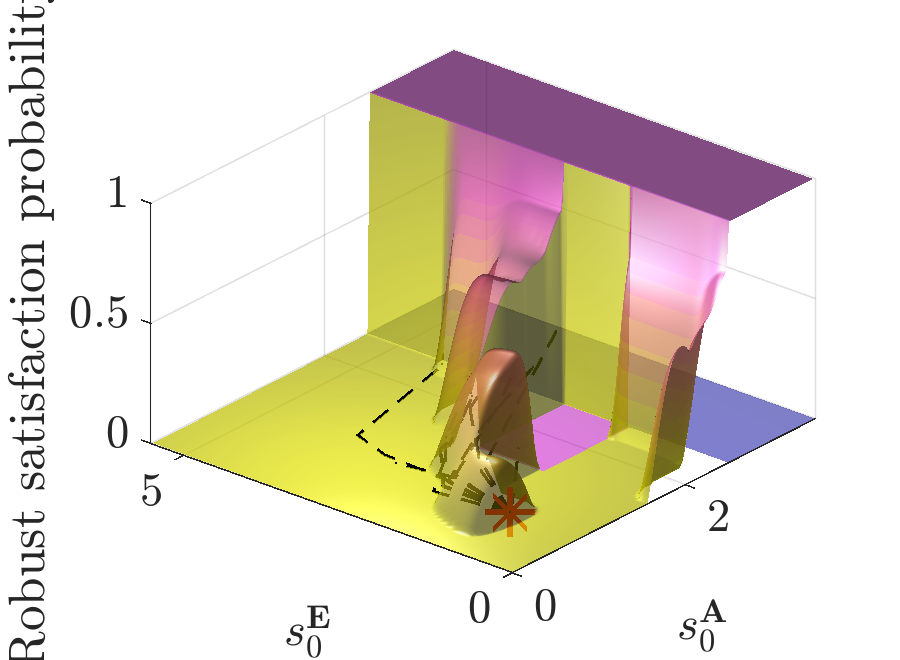}
    \caption{\mbox{$(v_0^\agent,v_0^\env)=(1.122,3.520)$} and $(s_0^\agent, s_0^\env)=\allowbreak(0.794,1.065)$}
    \end{subfigure}
    \caption{
    Robust satisfaction probability as a function of the initial position of $\agent\times\env$, for initial velocities
    $(v_0^\agent,v_0^\env)$.
    Additionally, 10 random trajectories (in black) starting from 
    $(s_0^\agent,s_0^\env)$ are shown, together with the collision $P_C$ (in magenta) and target $P_T$ (in blue) regions.}
    \label{fig:satProb_caseStudy}
\end{figure}

\paragraph{Abstraction-Based Control Design.} 
Throughout the running example, we established a quantified relation between the nominal model $\smash{(\agentred\times\simulator)(\hat\theta,\hat\adv)}$ and the concrete system $\smash{(\agent\times\env)(\theta)}$.
The following steps are performed using the toolbox SySCoRe \cite{van2023syscore}.
We discretize the state and input spaces $\smash{\Xh}$ and $\smash{\Uh}$ of $\smash{(\agentred\times\simulator)(\hat\theta,\hat\adv)}$ (into $327\times 10^6$ and $5$ partitions, respectively) to obtain a finite gMDP $(\agentred\times\simulator)_f$
and establish a simulation relation $\smash{\simrel{(\agentred\times\simulator)_f}{(\agentred\times\simulator)(\hat\theta,\hat\adv)}{\varepsilon_3}{\delta_3}}$ with $\varepsilon_3=0.2$ and $\delta_3=14.96\%$. 
Note that the magnitude of the error bounds $(\varepsilon_3,\delta_3)$ is linked directly to the precision of the chosen discretization, which is an inherent aspect of abstraction-based methods.
This underscores the importance of the new proxy framework, as it enables the computation of SSRs for high-dimensional systems.
The values of $(\varepsilon_3,\delta_3)$ can be reduced by refining the abstraction or adopting an adaptive gridding strategy. As the development of such techniques is orthogonal to our work, we use an equidistant gridding to simplify the presentation.
We perform controller synthesis on $\smash{(\agentred\times\simulator)_f}$ with $\varepsilon:=\smash{\sum_{i=1}^3\varepsilon_i}=0.2$ and $\delta:=\smash{\sum_{i=1}^3\delta_i}$ (max. $18.60\%$) using the transitivity property of $(\varepsilon,\delta)$-stochastic simulation relations \cite[Theorem~3]{schon2024bayesian} to obtain a robust controller $\C$ w.r.t. the specification $\psi$.
The corresponding robust satisfaction probability provided by SySCoRe is shown in Fig.~\ref{fig:satProb_caseStudy} as a function of the initial state of $\agent\times\env$.

\paragraph{Performance.}
To compare the performance of $\C$, we simulate the system $(\C\times\agent)\times\env$ from two different initial states.
Note that the specification includes an implicit time horizon, as the safety requirement $P_S$ mandates the systems to remain within the bounded state domain till $P_T$ is reached. Thus, a policy enforcing $\agent$ to wait for $\env$ to pass may not always be optimal.
We select the initial state
$x_0=(2.960,1.387,\allowbreak2.296,0.014)$ 
and run the system 1000 times. 
During runtime, it is verified whether the state of the environment $\env$ remains within the RoV $\X^\env$, rendering the simulator $\simulator$ a valid representation according to $\simreltwo{\simulator(\advAmbSet)}{\env}{}{}$.
Whence $x_{t}^\env\not\in\X^\env$ for some $t\in\N$, the AGC is violated.
We then stop the corresponding run and discard it.
As shown in Fig.~\ref{fig:satProb_caseStudy}(a), under the control policy obtained for the selected $x_0$, $\agent$ crosses the intersection \emph{before} $\env$.
The associated robust satisfaction probability obtained via SySCoRe is $p=58\%$, whereas the average experimental performance indicates an actual satisfaction probability closer to $p_{\text{real}}\approx67\%$.
We repeat the experiment for an initial state 
$x_0=(1.122,0.794,\allowbreak3.520,1.065)$,
yielding a policy of waiting (see Fig.~\ref{fig:satProb_caseStudy}(b)).
Due to the long time horizon, the robust satisfaction probability is low ($p=19\%$), whereas the average experimental performance indicates an actual satisfaction probability closer to $p_{\text{real}}\approx41\%$. 
This illustrates the conservatism introduced in bridging the gap between the surrogate and the concrete system
(Fig.~\ref{fig:idea}).

\section{Conclusion}\label{sec:conclusion}
We presented an assume-guarantee approach by establishing an adversarial probabilistic simulation relation between the surrogate and concrete closed-loop systems without directly computing concrete‑system error parameters.
By obviating domain discretization altogether, our approach enables scalable robust control synthesis for agents operating under environmental uncertainty.
By extension, this work takes a decisive step toward scalable solutions for partially observed systems.
We demonstrated its efficacy on a high‑dimensional nonlinear case study featuring tight agent–environment interactions.

\begin{credits}
\subsubsection{\ackname} 
This work is supported by the following grants: EIC 101070802 and ERC 101089047.
\end{credits}
%
%
%
\bibliographystyle{splncs04}
\bibliography{root}

\iflong
\appendix

\section{Proof of Theorem~\ref{thm:mainthm}}\label{app:proof_proxy_theorem}
\begin{proof}
We prove Theorem~\ref{thm:mainthm} by showing that the implication \eqref{eq:inductionmain_precond}~$\Rightarrow$~\eqref{eq:inductionmain_postcond} holds for all SSR conditions outlined in Definition~\ref{def:ssr}.
\paragraph{Condition (a).}
    According to Definition~\ref{def:ssr}(a), we have from the antecedent \eqref{eq:inductionmain_precond} with relation $\rel_a$ that
    \begin{equation*}
        \forall (x^\agent_0, x^\simulator_0)\!\in\!\X^\agent_0\times\X^\simulator_0,\,\exists (\hat x^{\agentred}_0,\hat x^\simulator_0)\!\in\!\X^{\agentred}_0\times\X^\simulator_0\!:\begin{bmatrix}\hat x^{\agentred}_0\\\hat x^\simulator_0\end{bmatrix}\!\rel\!\begin{bmatrix}F(x^\agent_0)\\x^\simulator_0\end{bmatrix}\leq\varepsilon.
    \end{equation*}
    From $\simreltwo{\simulator(\advAmbSet)}{\env}{}{}$, we have $\forall x^\env_0\in\X^\env_0,\,\exists{x}^\simulator_0\in\X^\simulator_0:\,x^\simulator_{0}=P{(x}^\env_{0})$ (Definition~\ref{def:BI}(a)), implying that
    \begin{equation*}
        \forall (x^\agent_0, x^\env_0)\!\in\!\X^\agent_0\times\X^\env_0,\,\exists (\hat x^{\agentred}_0,\hat x^\simulator_0)\!\in\!\X^{\agentred}_0\times\X^\simulator_0\!:\begin{bmatrix}\hat x^{\agentred}_0\\\hat x^\simulator_0\end{bmatrix}\!\rel\!\begin{bmatrix}F(x^\agent_0)\\P(x^\env_0)\end{bmatrix}\leq\varepsilon,
    \end{equation*}
    showing that condition (a) with relation $\rel_b$ holds for the consequent \eqref{eq:inductionmain_postcond}.
    
\paragraph{Condition (b).}
    Let $\rel_c$ be the relation in $\simreltwo{\simulator(\advAmbSet)}{\env}{}{}$.
    Via $(x^\simulator,x^\env)\in\rel_c$, every tuple $((\hat x^{\agentred},\hat x^\simulator),(x^\agent, x^\env))\in\rel_b$ associates at least one tuple $((\hat x^{\agentred},\hat x^\simulator),(x^\agent, x^\simulator))\allowbreak\in\allowbreak\rel_a$.
    For each fixed pair of tuples as initial conditions we propagate the systems $(\agentred\times\simulator)(\adv)$ and $(\agent\times_a\simulator)(\adv)$ by one time step.
    Based on Definition~\ref{def:ssr}(b), the antecedent \eqref{eq:inductionmain_precond} states that for every $u\in\U$ and $\adv\in\advAmbSet$ there exists a sub-probability kernel $\Wt(d\xhp^{\agentred}\times d\xhp^{\simulator}\times d\xp^{\agent}\times d\xp^{\simulator}|z,x^\simulator\!,u,\adv(\hat d^\simulator| \hat x^\simulator,\hat x^{\agentred}),\adv(d^\simulator| x^\simulator,F(x^\agent)))$ with $z:=(\hat x^{\agentred},\hat x^\simulator,x^\agent)$, that acts as a sub-probability coupling of 
    \begin{align}
    \begin{split}
        &\Tr^{\agentred}(d\xhp^{\agentred}|\hat x^{\agentred},\hat x^\simulator,u)
        \;\Tr^\simulator(d\xhp^{\simulator}|\hat x^\simulator,\hat x^{\agentred},\adv(\hat d^\simulator| \hat x^\simulator,\hat x^{\agentred})) \quad\text{and}
        \\
        &\Tr^\agent(d\xp^{\agent}| x^\agent,x^\simulator,u)
        \;\Tr^\simulator(d\xp^{\simulator}| x^\simulator, F(x^\agent), \adv(d^\simulator| x^\simulator,F(x^\agent)))
    \end{split}\label{eq:temp}
    \end{align}
    over $\rel_a$ with $\delta$.
    Simultaneously, the BI $\simreltwo{\simulator(\advAmbSet)}{\env}{}{}$ [Definition~\ref{def:BI}(b)] implies that for $\env$ initialized at some $x^\env\in\X^\env$ and propagated via $\Tr^\env(\cdotx| x^\env,x^\agent,u):\borel{\X^\env}\rightarrow[0,1]$, 
    there exists an adversary $\adv^\ast\in\advAmbSet$ resolving $\simulator$ to a prob. measure $\Tr^\simulator(\cdotx| x^\simulator, F(x^\agent), \adv^\ast(d^\simulator| x^\simulator,F(x^\agent))):\borel{\X^\simulator}\rightarrow[0,1]$ that assigns the same probability mass to each $(Q'\times Q)\in\rel_c$, i.e., 
    such that $\Tr^\env(Q| x^\env,x^\agent,u)=\allowbreak\Tr^\simulator(Q'| x^\simulator, F(x^\agent), \allowbreak\adv^\ast(d^\simulator| x^\simulator,F(x^\agent)))$ for all $(Q'\times Q)\in\rel_c$.
    Clearly, this witnessing measure $\Tr^\simulator(\cdotx|\ldots,\adv^\ast)$ (resolved by the witnessing adversary $\adv^\ast$) is included in the sub-probability coupling given between the measures in \eqref{eq:temp}.
    This implies that the induced sub-probability measure 
    \begin{equation*}
        \int_{\X^\simulator}\!\!\delta_{\xp^\simulator}(P( d\xp^\env)) \; \Wt(dz_+
        \times d\xp^{\simulator}
        \mid z,x^\simulator\!,u,\adv^\ast\!(\hat d^\simulator\mid \hat x^\simulator\!,\hat x^{\agentred}),\adv^\ast\!(d^\simulator\mid x^\simulator\!,F(x^\agent))),
    \end{equation*}
    for any $x^\simulator\in\X^\simulator$ s.t. $(x^\simulator,x^\env)\in\rel_c$ is a sub-probability coupling of
    \begin{align*}
        &\Tr^{\agentred}(d\xhp^{\agentred}\mid\hat x^{\agentred},\hat x^\simulator,u)
        \;\Tr^\simulator(d\xhp^{\simulator}\mid\hat x^\simulator,\hat x^{\agentred},\adv^\ast(\hat d^\simulator\mid \hat x^\simulator,\hat x^{\agentred}))\quad\text{and}
        \\
        &\Tr^\agent(d\xp^{\agent}\mid x^\agent,x^\simulator,u)
        \cdot
        \Tr^\env(d\xp^\env\mid x^\env,x^\agent,u)
    \end{align*}
    over $\rel_b$ with $\delta$ as before.

\paragraph{Condition (c).}
    Similar to case (a), according to Definition~\ref{def:ssr}(c), we have from the antecedent \eqref{eq:inductionmain_precond} with $\rel_a$ that
    \begin{equation*}
        \forall\!\left((\hat x^{\agentred},\hat x^\simulator),(x^\agent,x^\simulator)\right)\!\in\!\rel_a\!:\quad\mathbf{d}_{\Y^\agent\!\times\!\Y^\env}\!\left(\begin{bmatrix}h^\agent(\hat x^{\agentred})\\h^\env(\hat x^\simulator)\end{bmatrix}\!,\begin{bmatrix}h^\agent(F(x^\agent))\\h^\env(x^\simulator)\end{bmatrix}\right)\!\leq\!\varepsilon,
    \end{equation*}
    which implies condition (c) with $\rel_b$ for the consequent \eqref{eq:inductionmain_postcond} via $\simreltwo{\simulator(\advAmbSet)}{\env}{}{}$ (Definition~\ref{def:BI}(c)).\qed
\end{proof}

\section{MOR Coupling Compensator}\label{app:MOR_coupling_compensator}
Existing MOR techniques for stochastic simulation relations are scarce and mainly limited to linear dynamics~\cite{van2023syscore,wang2025unraveling}. Thus, they require additional steps, such as an \emph{piecewise affine} (PWA) abstraction layer, to address nonlinear systems. The latter has only ever been demonstrated on a 2D system (requiring several hours of runtime using parallel computing)~\cite{van2023syscore}. We consider this approach infeasible for the 8D system presented in the case study (see Section~\ref{sec:experiments}), whereas we show that our method establishes this model-based setting for the optimal $\delta_1 = \varepsilon_1 = 0$.

Beyond this, in this section, we compare the partial MOR coupling compensator from Proposition~\ref{prop:MOR_compensator_partial} to previous results.
Consider, for instance, the compensator for linear systems, proved in \cite[Theorem~8]{VanHuijgevoort2020SimQuant}, which can be extended into the following, more general form.
\begin{proposition}[MOR coupling compensator]\label{prop:MOR}
    Consider models
    \begin{align*}
        \M(\theta,\adv):&\left\{\begin{array}{ll}
            x_{t+1} &= f(x_t,u_t;\theta) + d_t + B_ww_t\\[.2em]
    	y_t &= h(x_t) \end{array}\right.
        ,&&\hspace{-1.5em}\begin{array}{ll}w_t&\sim \mathcal{N}_{\W}(\cdotx|0,I_n)\\[.2em]
        d_t&\sim\adv(\cdotx| x_t)\end{array},\\
        \M_r(\theta,\adv_r):&\left\{\begin{array}{ll}
            x_{r,t+1} &= f_r(x_{r,t},u_{r,t};\theta) + d_{r,t} + B_{w,r}w_{r,t}\\[.2em]
    	y_{r,t} &= h_r(x_{r,t}) \end{array}\right.
        ,&&\hspace{-1.5em}\begin{array}{ll}w_{r,t}&\sim \mathcal{N}_{\W}(\cdotx|0,I_n)\\[.2em]
        d_{r,t}&\sim\adv_r(\cdotx| x_{r,t})\end{array},
    \end{align*}
    with a matrix $F'\in\R^{n\times n_r}$ such that $F'x_r\mapsto x$, noise coupling $\gamma_t:=w_t- \new{w_{r,t}}\in\Gamma$, and interface function $u_t\sim\InFu(\cdotx| x_t, x_{r,t}, u_{r,t})$,
    leading to error dynamics
    \begin{equation*}
        \Delta\xp = \Delta_r(x,x_r,u_r;\theta) + B_w\gamma + (B_w-F'B_{w,r})w_r,
    \end{equation*}
    with an offset given by
    \begin{equation*}
        \Delta_r(\cdotx):=f(x,\InFu(u|\cdotx);\theta)+\adv(d|x) - F'\left( f_r(x_r,u_r;\theta)+\adv_r(d_r|x_r)\right).
    \end{equation*}
    For a bounded set $\Gamma$, define a \emph{controlled-invariant}\footnote{A set $\rel_r'$ is controlled-invariant for a bounded set $\Gamma$, if we have that $\forall (x_r,x)\in\rel_r',\,\exists\gamma\in\Gamma:\,\Delta\xp\in\rel_r'$ for all $(u_r,w_r)\in\U_r\times\W$.
    \new{If $\W$ is unbounded, the computations are performed for a truncated $w_r\in\W_{\text{trunc}}\subset\W$ that captures $1-\delta_{\text{trunc}}$ of the probability. Then, the statement holds for $\delta_r\geq\delta_{\text{trunc}}+\sup_{\gamma\in\Gamma}1-2\cdf{-\frac{1}{2}\norm{\gamma}}$.}
    } set
    \begin{equation*}
        \rel_r':=\left\lbrace(x_r,x)\in\X_r\times\X\mid\norm{x-F'x_r}_{D_r}\leq\varepsilon_r\right\rbrace.
    \end{equation*}
    If there exist $\varepsilon_r\geq0$ and $\delta_r\in[0,1]$ such that
    \begin{itemize}
        \item $\varepsilon_r\geq\sup_{(x_r,x)\in\rel_r'}\norm{h(x)-h_r(x_r)}$; and
        \item $\delta_r\geq\sup_{\gamma\in\Gamma}1-2\cdf{-\frac{1}{2}\norm{\gamma}}$;
    \end{itemize}
    then, $\simrel{\M_r(\theta,\adv_r)}{\M(\theta,\adv)}{\varepsilon_r}{\delta_r}$.
\end{proposition}
Note, that for $\varepsilon_r=0$ this setup requires a matrix $F'$ that associates every state $x_r\in\X_r$ with a \emph{unique} matching state $x\in\X$.
This can be very restrictive. We may not be able/willing to construct a reduced-order model $\M_r$ that can match every state $x$ with a concrete counterpart $x_r$ such that $x = F'x_r$, i.e., condensing all information of $x$ into a lower-dimensional $x_r$.
For $\varepsilon_r>0$, we must propagate the set $\rel_r'\subset\X_r\times\X$ through the (potentially high-dimensional) dynamics of $\M_r\times\M$ to show controlled invariance.
In our case study (Section~\ref{sec:experiments}), this would require computations on the original 8D system to establish the relation.
Instead, Proposition~\ref{prop:MOR_compensator_partial} allows us to relate models with states of asymmetric information content without the need to consider truncated dimensions of $\M$, lifting prior requirement.
The arising ambiguity in state pairings $(x_r,x)\in\rel_r$ remains inherent in the relation $\rel_r$ (Proposition~\ref{prop:MOR_compensator_partial}), thus, we call this a \emph{partial} coupling compensator.

\section{Case Study Details}\label{app:case_study}
The specification is defined based on the regions for collision $P_C:=[0,3]\times[1.5,2.5]\times[0.75,3.75]\times[1.5,2.5]$, target $P_T:=[0,3]\times[2.5,3.5]\times[0.75,3.75]\times[0,5.5]$, and safety $P_S:=\X^\agent\times\X^\env$.

\subsection{Agent Dynamics}\label{app:agent_dynamics}
We provide further details on the dynamics of the agent vehicle $\agent$ in \eqref{eq:agent_car}. 
The following spaces are used: The state space $\X^\agent:=\X^\agent_\xi\times\X^\agent_v\times\X^\agent_s:=[-0.05,0.05]\times[0,3]\times[0,3.5]$, initial set $\X^\agent_0=\X^\agent$, input space $\U^\agent:=[-5,5]$, and output space $\Y^\agent:=\X^\agent_v\times\X^\agent_s$.
The noise distributions are
$p^\agent_{w,\xi}(dw^\agent_\xi):=\mathcal{N}(dw^\agent_\xi|0,10^{-3})$, %
$p^\agent_{w,v}(dw^\agent_v):=\mathcal{N}(dw^\agent_v|0,0.2)$, and
$p^\agent_{w,s}(dw^\agent_s):=\mathcal{N}(dw^\agent_s|0,0.1)$.

\subsection{Environment Dynamics}\label{app:env_dynamics}
Here, we provide additional details on the dynamics of the environment vehicle $\env$ in \eqref{eq:env_car}.
The following spaces are used: The state space $\X^\env:=[-0.05,0.05]^3\times[0.75,3.75]\times[0,5.5]\times[-0.5,0.5]$, initial set $\X^\env_0=\X^\env$, and output space $\Y^\env:=[0.75,3.75]\times[0,5.5]$.
The noise is distributed according to $p^\env_w(dw^\env):=\mathcal{N}_{\W^\env}(dw^\env|\allowbreak0,\allowbreak\mathrm{diag}(0.01,0.01,\allowbreak0.01,0.2,0.1,0.01))$, truncated to $\W^\env:=[-0.05,0.05]^3\allowbreak\times[-10^3,\allowbreak10^3]\allowbreak\times[-10^3,10^3]\allowbreak\times[-0.5,0.5]$.
The derivative terms are as follows:%
\begin{equation*}
    \dot\beta(\beta,\mathrm{d}\Psi,v) := \frac{-Cg\mu\beta}{v}
    -\mathrm{d}\Psi, \quad
    \ddot\Psi(\beta,\mathrm{d}\Psi,v) := \frac{-Cgl\mu m\mathrm{d}\Psi}{I_zv},
\end{equation*}
with parameters $C:=1$, $g:=9.81$, $\mu:=0.9$, $l:=2.579$, $m:=1093$, and $I_z:=1792$.
%

\subsection{Reduced-Order Agent}\label{app:reduced_agent}
For the reduced agent $\agentred$ in \eqref{eq:agent_car_reduced}, we select the spaces $\X^{\agentred}_{v}:=[0,3]$, $\X^{\agentred}_{s}:=[0,3.5]$, and $\Y^{\agentred}:=\Y^\agent$ --- matching those of the full agent $\agent$.

We move on with establishing the relation $\simrel{(\agentred\times\simulator)(\theta,\adv)}{(\agent\times_a\simulator)(\theta,\adv)}{{\varepsilon_1}}{\delta_1}{}{}$, $(\theta,\adv)\in\Theta\times\advAmbSet$.
For this, consider the two composed systems $\M:=(\agent\times_a\simulator)(\theta,\adv):=(\X^\agent\times\X^\simulator,\X^\agent_0\times\X^\simulator_0,\U,\D,\Tr,h,\Y)$ and $\M_r:=(\agentred\times\simulator)(\theta,\adv):=(\X^{\agentred}\times\X^\simulator,\X^{\agentred}_0\times\X^\simulator_0,\U_r,\D_r,\Tr_r,h_r,\Y)$, mapping $\times_{a\!}:(x^{\agent},x^\simulator)\mapsto(F(x^{\agent}),x^\simulator)$ with $F:x^{\agent}\mapsto[0_{2\times1},I_2]x^{\agent}$, interface function $u= u_r$, and $\varepsilon_1=0$ which implies the relation $\rel_r=\{(x_r,x)\in(\X^{\agentred}\times\X^\simulator)\times(\X^\agent\times\X^\simulator)| \times_{a\!}(x)=x_r\}$.
It follows immediately that $\sup_{(x_r,x)\in\rel_r}\norm{\times_{a}(x)-x_r}=0\leq\varepsilon_1$.

A sub-probability coupling of $\Tr_r$ and $\Tr$ can be derived as follows.
Firstly, it is easy to verify that a sub-probability measure 
\begin{equation*}
    \Wt(Q_r\times Q|\cdotx) \!:=\! \min\!\left( \Tr_r(Q_r|x_r,u_r,\adv(d_r|x_r);\theta), \Tr(Q|x,u_r,\adv(d|x_r);\theta) \right)\\
\end{equation*} 
specifies such a sub-probability coupling.
Next, we find $\delta_1$ such that $\Wt(\rel_r)\geq 1-\delta_1$.
For this, we define the sets $\W:=\W^\agent\times\W^\simulator$, $\W_r:=\W^{\agentred}\times\W^\simulator$, and matrices $\Sigma_r := \mathrm{diag}(0.2,0.1,0.2,0.1)$, $\Sigma := \mathrm{diag}(\sigma_\xi^2,\Sigma_r)$, with $\sigma_\xi^2:=10^{-3}$, and
\begin{align*}
    A_r &:= \begin{bmatrix}
        \theta^\agent&0&0&0\\
        \tau&1&0&0\\
        0&0&\theta^\env&0\\
        0&0&0&1
    \end{bmatrix},
    && B_r :=\begin{bmatrix}
        \tau\\
        0\\
        0\\
        0
    \end{bmatrix},
    && N_r := \begin{bmatrix}
        0&0\\
        0&0\\
        1&0\\
        0&1
    \end{bmatrix},\\
    A &:= \left[\begin{array}{l|c}
        a_1 & \begin{matrix}0&\cdots&0\end{matrix}\\\hline
        \begin{matrix}a_2\\a_3\\a_4\\a_5\end{matrix} & A_r
    \end{array}\right],
    && B :=\begin{bmatrix}
        b\\\hline
        B_r
    \end{bmatrix},
    && N := \left[\begin{array}{c}
        \begin{matrix}0&0\end{matrix}\\\hline
        N_r
    \end{array}\right].
\end{align*}
For any $z:=(x_r,x,u_r)\in\rel_r\times\U_r$, we get via $\mathcal{N}_\W(dw_\xi\times dw_r|0,\Sigma)\allowbreak\equiv\allowbreak\mathcal{N}_{\W_\xi^\agent}(dw_\xi|\allowbreak0,\sigma_\xi^2)\allowbreak\;\mathcal{N}_{\W_r}(dw_r|0,\Sigma_r)$, $\W\equiv\W_\xi^\agent\times\W_r$, and $Q\equiv Q_\xi\times Q_r'$:
\begin{align}
    &\Wt(\rel_r|z) \!=\!\!\! 
    \int\!\!\!\!\int\!\!\!\!\int_{{\begin{array}{l}
        dw_r\in\W_r\\
        dw_\xi\in\W_\xi^\agent\\
        (Q_r\times (Q_\xi\times Q_r'))\in\rel_r
    \end{array}}}\hspace{-6em}
    \min\!\Big(
    \delta_{w_r}(Q_r),\delta_{[a_2,a_3,a_4,a_5]\T\xi^\agent+w_r}(Q_r')\delta_{a_1\xi^\agent+bu_r+w_\xi}(Q_\xi)
    \Big)\nonumber\\
    &\hspace{10em}
    \mathcal{N}_{\W_\xi^\agent}(dw_\xi|0,\sigma_\xi^2)\mathcal{N}_{\W_r}(dw_r|\mu,\Sigma_r),\label{eq:temp_wt}
\end{align}
where we rebased via $\hat w:=\mu+w$, with $\mu(x_r,u_r):=A_rx_r+B_ru_r+N_r\adv(x_r)$.
Recall that $\xi^\agent:=[1,0,0,0,0]x$.
Here,
we consider the worst-case $\xi^\agent\in\X_\xi^\agent$.

For the expression \eqref{eq:temp_wt} in its general form, $\delta_1$ can be determined via Monte Carlo approximation.
We choose $a_2=a_3=a_4=a_5=0$, thus, the expression reduces due to $\forall (Q_r,(Q_\xi\times Q_r'))\in\rel_r:\,Q_r=Q_r'$ and $Q_\xi\equiv\X_\xi^\agent$ (by the definition of $\rel_r$) to
\begin{align*}
    \Wt(\rel_r|z) &= 
    \int_{Q_r\in\X_r} \int_{Q_\xi\in\X_\xi^\agent} \int_{\W_r} \int_{\W_\xi^\agent} 
    \min\Big(
    \delta_{w_r}(Q_r),\Big.\\
    &\qquad\Big.\delta_{w_r}(Q_r)\delta_{a_1\xi^\agent+bu_r}(Q_\xi)
    \Big)
    \mathcal{N}_{\W_\xi^\agent}(dw_\xi|0,\sigma_\xi^2)\mathcal{N}_{\W_r}(dw_r|\mu,\Sigma_r),\\
    &= 
    \int_{Q_\xi\in\X_\xi^\agent} 
    \mathcal{N}_{\W_\xi^\agent}(Q_\xi|a_1\xi^\agent+bu_r,\sigma_\xi^2).
\end{align*} 
Since $\W_\xi^\agent=\R$, we obtain $\delta_1$ as a function of $u_r$, namely
\begin{equation*}
    \delta_1(u_r) \geq 1 - \inf_{\xi^\agent\in\X_{\xi}^\agent} \int_{Q_\xi\in(\X_\xi^\agent\ominus a_1\xi^\agent)} 
    \mathcal{N}(Q_\xi|bu_r,\sigma_\xi^2).
\end{equation*}
For $a_1=0.2$ and $b=0.001$, we have $\delta_1(u_r)=0\%$.

\subsection{Ambiguity Compensation}\label{app:ambiguity_compensation}
Consider the setup in Theorem~\ref{thm:ambiguity_compensation}. For the systems given in the case study, the error dynamics \eqref{eq:Delta} reduce to
\begin{equation*}
    \Delta(\hat x;\hat\theta,\theta,\beta,\Psi) 
    \!=\!
    (\theta^\agent-\hat\theta^\agent)\hat v^\agent + (\theta^\env-\hat\theta^\env)\hat v^\simulator + \tau \hat v^\simulator(\cos(\beta+\Psi)-1),
\end{equation*}
where $\hat x:=[\hat s^{\agent}, \hat v^{\agent}, \hat s^{\simulator}, \hat v^{\simulator}]\T$, $\beta\in\X_\beta^\env$, $\Psi\in\X_\Psi^\env$, and $\varepsilon_2=0$.
From \eqref{eq:delta}, we have 
\begin{equation*}
        \delta_2(\hat x;\hat\theta) = 1-2\,\cdf{-\frac{1}{2}
        \sup_{\theta\in\Theta,\beta\in\X_\beta^\env,\, \Psi\in\X_\Psi^\env}
        \norm{\Delta(\hat x;\hat\theta,\theta,\beta,\Psi)}}.
\end{equation*}
We visualize $\delta_2(\hat x;\hat\theta)$ in Fig.~\ref{fig:delta_ambiguity} as a function of $(\hat v^{\agent}, \hat v^{\simulator})$.
\begin{SCfigure}
    \includegraphics[width=.7\linewidth]{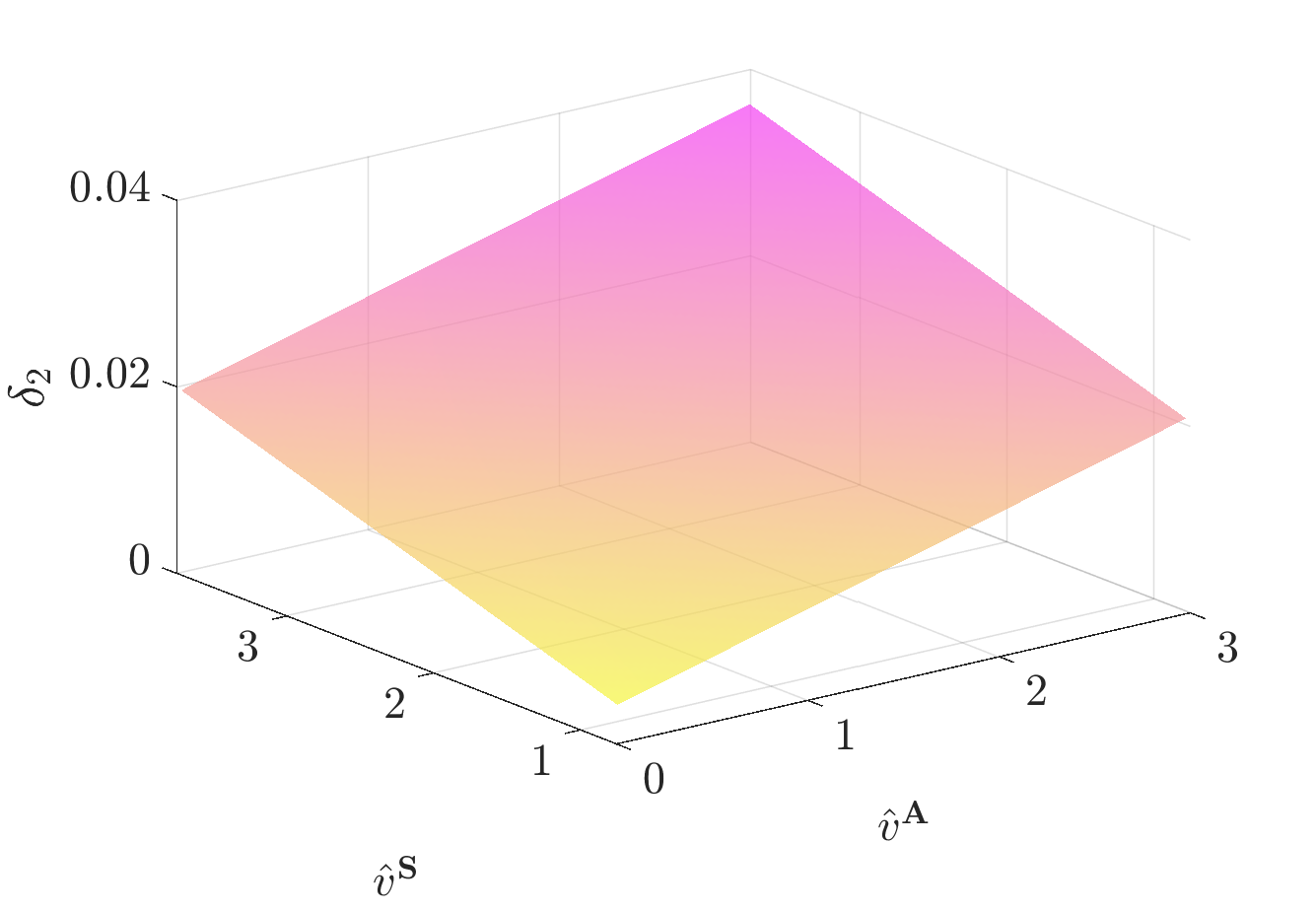}
    \caption{Value of $\delta_2$ as a function of the velocities $(\hat v^{\agent}, \hat v^{\simulator})$.}
    \label{fig:delta_ambiguity}
\end{SCfigure}
\fi

\end{document}